\documentclass[aps,physrev,twocolumn,superscriptaddress]{revtex4-1}
\usepackage{amsmath}
\usepackage{graphicx}
\usepackage{amssymb}
\usepackage{rotating}
\usepackage{tabularx}
\usepackage{booktabs}
\usepackage{bm}
\usepackage{comment}
\usepackage{array}
\usepackage{enumitem}
\usepackage{multirow,makecell}
\usepackage{xcolor}
\usepackage{colortbl}
\usepackage{tikz}
\usetikzlibrary{matrix}
\usetikzlibrary{shapes,arrows}
\usepackage{orcidlink}
\usepackage[]{hyperref}
\usepackage[absolute,overlay]{textpos}

\newcommand{\nocontentsline}[3]{}
\let\origcontentsline\addcontentsline
\newcommand\stoptoc{\let\addcontentsline\nocontentsline}
\newcommand\resumetoc{\let\addcontentsline\origcontentsline}
\newcommand{\pDrg}{pN$\cdot$nm$\cdot$s/$\rm{rad}^{2}$}

\hypersetup{
	colorlinks=false,
	linkcolor=black,
	pdfborder={0 0 0},
	urlcolor=black,
	citecolor=black
}

\begin{document}

\title{Reconstruction of the Bacterial Flagellar Motor's Energy Landscape, Viscous Load, and Torque Generation Across Diffusion Regimes}

\author{N. J. L\'opez-Alamilla}\email[]{n.jared.lopezalamilla@gmail.com} 
\affiliation{Laboratoire Charles Coulomb (L2C), Univ. Montpellier, CNRS, Montpellier, France}
\author{A. L. Nord}
\affiliation{Centre de Biologie Structurale (CBS), Univ. Montpellier, CNRS, INSERM, Montpellier, France}
\author{F. Pedaci}
\affiliation{Centre de Biologie Structurale (CBS), Univ. Montpellier, CNRS, INSERM, Montpellier, France}
\author{J. Palmeri} 
\affiliation{Laboratoire Charles Coulomb (L2C), Univ. Montpellier, CNRS, Montpellier, France}
\author{N.-O. Walliser}\email[]{nils-ole.walliser@umontpellier.fr} 
\affiliation{Laboratoire Charles Coulomb (L2C), Univ. Montpellier, CNRS, Montpellier, France}

\date{\today}

\begin{abstract}
	The bacterial flagellar motor (BFM) converts transmembrane ion flux into directed mechanical rotation, driving bacterial motility. 
	Despite extensive study, the frictional forces and energetics governing its torque generation remain poorly understood.
	Here, we combine single-molecule rotation measurements with stochastic thermodynamics to quantitatively estimate its effective torque, viscous drag and activation energy barriers. 
	We present three complementary methods based on solutions to the Smoluchowski equation for overdamped diffusion in a tilted periodic potential, which use as input the steady-state angular velocity and rotational diffusion data from individual \textit{E. coli} motors spanning different dynamical regimes. 
	Crucially, these three methods require neither active external torque control, nor prior knowledge of the system’s viscous drag or the motor’s torque output.
	The first method assumes as input a model-dependent sinusoidal potential (single Fourier mode), albeit with unknown periodicity, yielding closed-form results in the low- and high-tilt limits, whereas the last two methods use a full Fourier-based reconstruction to output the now model-independent potential landscape.
	These approaches yield consistent estimates of the potential periodicity 
	($\approx$ 26-fold symmetry), energy barrier height ($\approx 2\!-\!4 k_{\rm B}T$), and  internal friction coefficient ($\approx 0.1$ \pDrg). 
	Our results reveal that the BFM's torque-velocity relationship deviates significantly
	from the linear approximation near the critical tilt, where angular diffusion is maximized. 
	More broadly, our framework provides a coherent strategy for reconstructing nanoscale energy landscapes from single-molecule data and is generalizable to other stepping molecular motors operating in cyclic conditions.
\end{abstract}

\maketitle

\tableofcontents

\resumetoc
\section{Introduction}
Molecular motors are self-assembling biological machines composed of proteins that are able to convert (electro-)chemical energy into mechanical work. They play essential roles in processes such as intracellular transport, cellular motility, muscle contraction, and ATP synthesis\,\cite{Alberts2014, Howard2001}. Their operation is governed by underlying energy landscapes that dictate how motion proceeds through cycles of binding, catalysis, and conformational change. For rotary motors such as ATP synthase and the bacterial flagellar motor (BFM), the energy landscape is periodic in the angular coordinate. When combined with a sustained driving force, such as ATP hydrolysis or ion translocation down an electrochemical gradient, the system can be modeled as overdamped Brownian motion over a tilted periodic potential\,\cite{Reimann2002, Astumian1997}.

The BFM is the rotary molecular motor embedded in the bacterial cell envelope that drives the rotation of the extracellular helical flagellum, enabling swimming, chemotaxis, and infection \,\cite{BergAnderson1973,MinaminoImada2015}. Structurally 
(see Figs.\,\ref{Fig:1}a and b), the motor comprises multiple protein rings embedded in the membranes and cytoplasm, with the cytoplasmic portion forming the rotor\,\cite{MinaminoImada2015,Johnson2021}. A central rod extends from the rotor through the peptidoglycan (cell wall) and outer membrane, where its high-speed rotation is supported by the LP ring, a pair of protein rings embedded in the peptidoglycan and outer membrane that function together as a molecular bearing\,\cite{Yamaguchi2021, Johnson2021}. The motor is powered by the flow of ions down an electrochemical gradient across the inner membrane; the ions pass through stator units anchored to the peptidoglycan at the rotor's periphery, which convert electrochemical energy into mechanical torque\,\cite{KojimaBlair2004}.

Single-molecule experiments have shown that the BFM rotates in approximately 26 discrete steps per revolution\,\cite{SowaRoweLeakeYakushiHommaIshijimaBerry2005, Nakamura2010, JohnsonFurlongDemeNordCaesarChevanceBerryHughesLea2021, Drobnic2025}. These steps were initially interpreted as elementary torque-generating events, attributed to periodic engagement between stator units and the rotor protein FliG\,\cite{SowaRoweLeakeYakushiHommaIshijimaBerry2005}. However, subsequent structural studies revealed that the rotor contains 34 FliG subunits\,\cite{Tan2024}, while the LP ring is the only component of the motor with 26-fold symmetry 
(see Figs.\,\ref{Fig:1}a and b)\,\cite{Johnson2021}. This prompted a reinterpretation: the observed stepping arises not from torque generation, but from interactions between the rotating rod and the LP ring, which imposes a periodic energy landscape on its motion.

The LP ring represents a unique mechanical solution in biology, enabling stable high-speed rotation while likely minimizing mechanical resistance. Its function relies on electrostatic repulsion and steric alignment between the rod and the bearing\,\cite{Yamaguchi2021, Nakamura2023}, 
plausibly producing a 26-fold periodic potential that influences the motor’s torque dynamics \cite{Xing2006,MoraYuSowaWingreen2009}. In addition to stabilizing the rod, 
the LP ring introduces a friction-like landscape that shapes the effective energy profile experienced by the motor. Despite its central mechanical role, the energy landscape imposed by the LP ring remains largely unquantified \,\cite{MoraYuSowaWingreen2009}, with only recent preliminary estimates available\,\cite{RieuNordCourbetElSayyedEtAl2025}.

In recent work, Hayashi et al. \cite{HayashiSasakiNakamuraKudoInoueNoji2015} used electrorotation to apply a controlled external torque to F$_1$-ATPase while measuring changes in rotational velocity and diffusion. They observed that diffusion peaked near the critical tilt condition, where the applied torque balances the energy barriers in the motor’s periodic potential. This enabled direct reconstruction of the underlying energy landscape without assumptions about internal friction or kinetic details. Such experiments are possible for soluble motors like F$_1$-ATPase, where torque can be directly applied in reconstituted \textit{in vitro} assays. For the BFM, however, these measurements are far more challenging. Although optical \cite{Berry1997}, magnetic \cite{Wang2022}, or electrorotation techniques \cite{Berry1995} can in principle be used to impose torque, most experimental studies rely on passive observation of rotation, where neither the torque generated by the motor nor the rotational drag coefficient (including internal and probe contributions) are directly measurable. In typical assays, the only available observables are the steady-state angular velocity and the effective rotational diffusion coefficient, both extracted from the motion of a tethered bead. Given that the LP ring imposes a strong periodic potential that dominates the BFM dynamics in this regime, we ask whether these observables can be used to quantitatively reconstruct its energy profile.

Here, we develop a stochastic thermodynamics framework for rotational motion in a tilted periodic potential. Motivated by typical BFM experiments, we ask whether a simple combination of the measured angular velocity and diffusion can serve as a reliable proxy for the underlying energy landscape. We show that this is possible across three distinct regimes: low-tilt, where motion is barrier-limited; high-tilt, where free drift dominates; and an intermediate (critical-tilt) regime, where diffusion is maximized. These correspond to torque values that are small, large, or comparable to the height of the underlying energy barriers. Our approach enables the quantitative reconstruction of the LP ring’s energy landscape and the internal drag coefficient of the motor, even when torque and drag are unknown. More broadly, the same framework applies to any molecular motor operating in a periodic energy landscape, particularly in systems where active control is limited, forces cannot be measured directly, or potentially when the landscape is heterogeneous or time-dependent \cite{RieuNordCourbetElSayyedEtAl2025}. By providing a unified description of motor dynamics across physical regimes, this work offers a new theoretical  tool for probing energy transduction at the nanoscale.

\section{Single motor measurements}
\label{sec:SingleMotorMeasuremets}
We measured the steady-state rotation of 95 individual \textit{Escherichia coli} BFMs via a bead assay \cite{Hoffmann2025}, tracking the rotation of $100$ nm diameter gold beads attached to the flagellar hook
(see Fig.\,\ref{Fig:1}c), as described in Materials and Methods. The data were acquired at an acquisition rate of 109,500 frames per second. The bead center was tracked as it rotated slightly off-axis, yielding its circular $x(t), y(t)$ trajectory (see Fig.\,\ref{Fig:1}d), the BFM rotation angle, $\phi(t) = \arctan(y(t)/x(t))$, and the instantaneous angular speed, $\omega(t) = \dot{\phi}(t)$. From each motor, we extracted the steady-state speed, $\omega_{\rm ss} \simeq \langle \phi \rangle / \Delta t$, and the rotational diffusion coefficient, $D_{\rm ss} \simeq \left( \langle \phi^2 \rangle - \langle \phi \rangle^2 \right) / (2 \Delta t)$ from the longest portion of the trace exhibiting constant speed (see Materials and Methods). 

We analyzed the steady-state angular probability distribution $P_{\rm ss}(\phi)$ for each motor using Ensemble Empirical Mode Decomposition (see Materials and Methods) to extract the dominant periodic components. As shown in Fig.\,\ref{Fig:1}e, we found a global dominant periodicity of 25.7 $(\pm\,13)$ preferred positions per revolution, in agreement with previous studies that reported approximately 26\,\cite{SowaRoweLeakeYakushiHommaIshijimaBerry2005, Nakamura2010, JohnsonFurlongDemeNordCaesarChevanceBerryHughesLea2021, Drobnic2025}, and consistent with the 26-fold symmetry of the  LP ring.

\section{Background theory of diffusion in a periodic tilted potential}

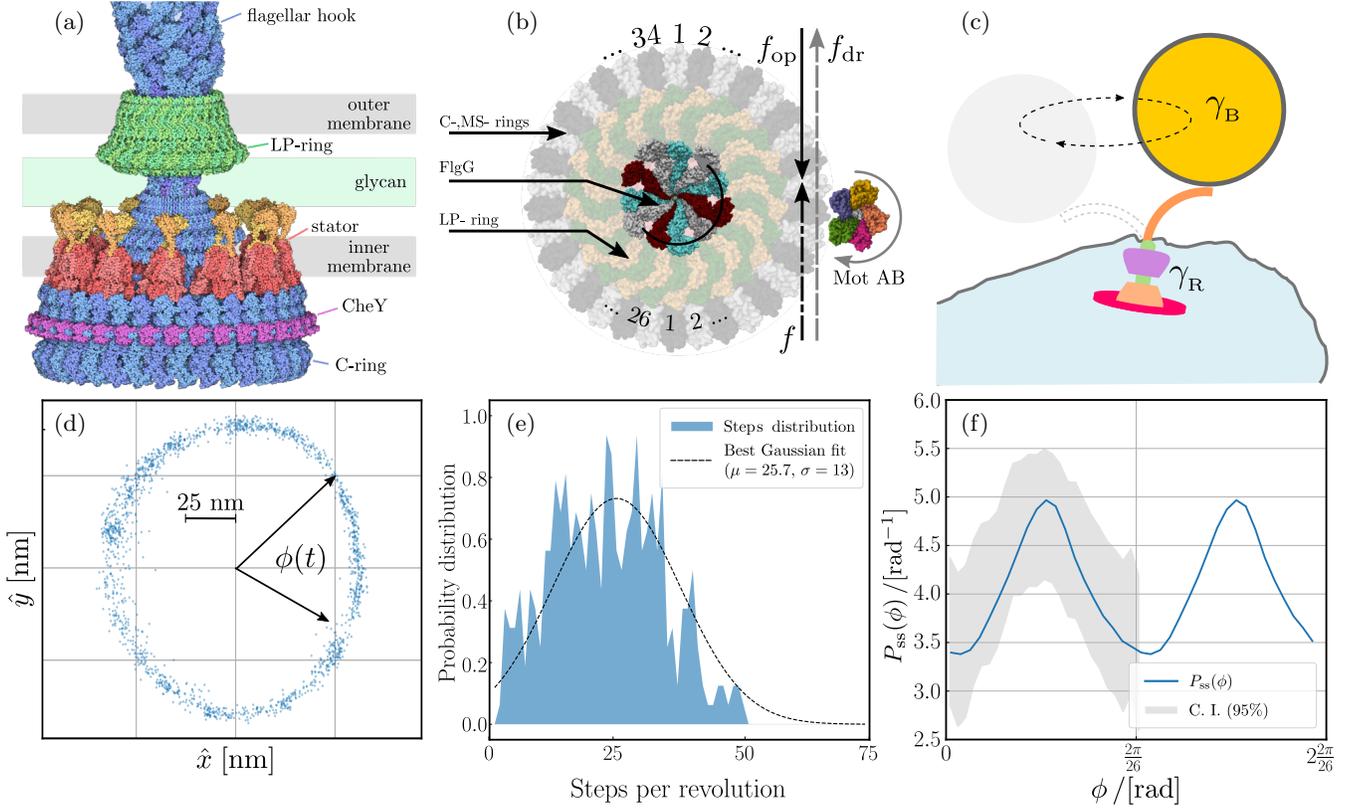
\begin{figure*}
	\includegraphics[width=0.99\textwidth]{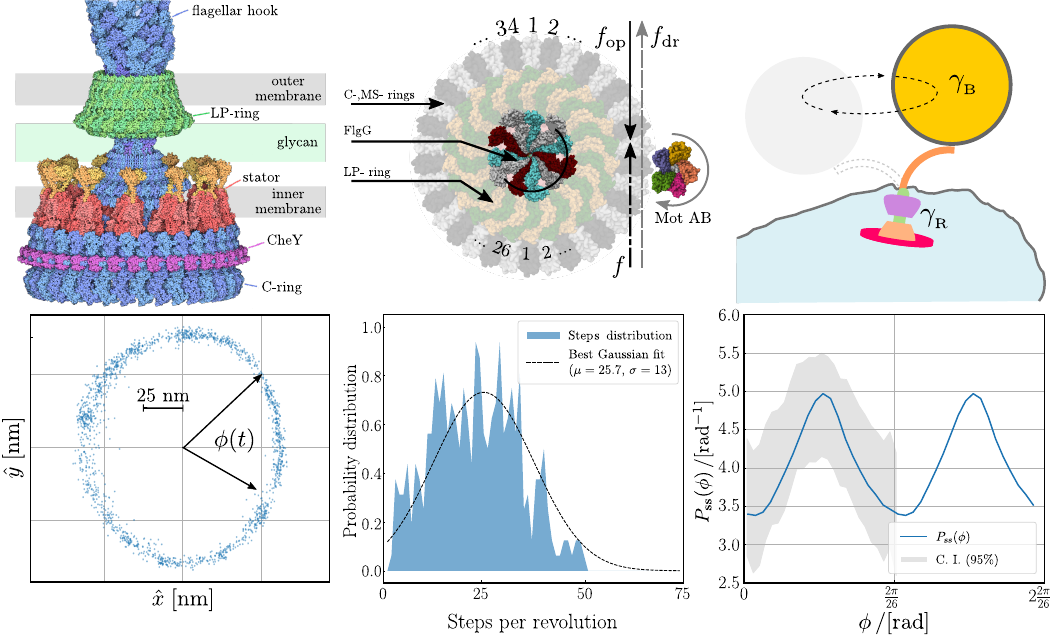}\put(-486,295){(a)}\put(-486,143){(d)}\put(-315,295){(b)}\put(-315,143){(e)}\put(-143,295){(c)}\put(-143,143){(f)}  
	\caption{ 
		a) Side view of the BFM's protein structure (by David Goodsell 10.2210/rcsb\_pdb/mom\_2024\_12 under  CC-BY-4.0 license). b) Composed top view displaying key folding periodicity of the BFM's components and a schematic representation of the torques involved 
		($f_{\rm dr}$, $f_{\rm op}$, $f$) in its motion. 
		c) Schematic representation of the BFM bead assay and the relevant viscous loads present, 
		$\gamma_{\textsc{b}}$ and $\gamma_{\textsc{r}}$, for the bead and internal ring, respectively. 
		d) Example of the angular position time series obtained from single-molecule experiments tracking the center of mass of the attached bead. 
		e) Ensemble Empirical Mode Decomposition (EEMD) analysis of the stepping size for a set of 95 individual traces showing a mean stepping very close to 26 steps per revolution. 
		f) Normalized probability distribution of the angular position folded over the main periodicity found by the EEMD analysis (C.I. denotes confidence interval).}
	\label{Fig:1}
\end{figure*}

Motivated by the periodicity observed in the angle $\phi(t)$, we aimed to develop analytical methods to extract features of the underlying energy landscape from the experimentally accessible quantities $P_{\rm ss}(\phi)$, $\omega_{\rm ss}$, and $D_{\rm ss}$.
Our approach builds on the established theoretical framework for one dimensional diffusion in tilted periodic potentials, a class of systems described by different formulations\,\cite{Risken1996, Gardiner2009}. In this work, we focus on the Smoluchowski formulation and present its key analytical results.

We write the potential as $V(\phi) = V_{\rm eq}(\phi) - f\phi$, where $\phi$ is the azimuthal angle of rotation, $f$ is the effective  constant tilt torque (resulting from the difference between driving and opposing torques on the system $f=f_{\rm dr}-f_{\rm op}$, as seen in Fig.\,\ref{Fig:1}b). Given the observed quasi single-mode periodicity, the equilibrium potential is chosen as $V_{\rm eq}(\phi) = \frac{E_{\rm a}}{2}\cos(2\pi \phi/L)$, presenting energy barriers of height $E_{\rm a}$, and periodicity $L$ (corresponding to a $\mathcal{P}$-fold symmetry with $\mathcal{P}=2\pi/L$). 
The evolution of the system is described by its probability density function $P(\phi,t)$, which can be computed using the Smoluchowski equation:
\begin{equation}\label{Eq:Evol/Smoluchowski/rSpace}
	\frac{\partial}{\partial t}P(\phi,t)=\mathcal{L}P(\phi,t)\,,
\end{equation}
where $\mathcal{L}$ is the evolution operator defined by
\begin{equation}
	\label{Eq:evol_operat}    
	\gamma\mathcal{L} = \frac{\partial}{\partial \phi}\left[\frac{\partial V(\phi)}{\partial \phi} + \Theta\frac{\partial}{\partial \phi}\right], 
\end{equation}
with $\gamma$ the viscous drag coefficient and $\Theta = k_{\rm B}T$ the thermal energy\,\cite{Gardiner2009}. 
We are interested in the steady state of the system where closed solutions for $P_{\rm ss}$, $\omega_{\rm ss}$, and $D_{\rm ss}$ are known\,\cite{ReimannVandenBroeckLinkeEtAl2001} and can be written as 
\begin{equation}\label{Eq:Pss/Reimann}
	P_{\rm ss}(\phi)=\frac{I_+(\phi)}{\left\langle I_+(\phi)\right\rangle_{_L}}\,,
\end{equation}
\begin{equation}\label{Eq:vss/Reimann}
	\omega_{\rm ss}= \lim_{\Delta t\rightarrow \infty}\frac{\langle \phi\rangle}{\Delta t}=\frac{L\Theta}{\gamma}\frac{\left(1-e^{-fL/\Theta}\right)}{\left\langle I_+(\phi)\right\rangle_{_L}}\,,
\end{equation}
\begin{equation}\label{Eq:Dss/Reimann}
	D_{\rm ss}= \lim_{\Delta t\rightarrow \infty}\frac{\langle \phi^2\rangle-\langle \phi\rangle^2}{2\Delta t}=\frac{L^2\Theta}{\gamma}\cfrac{\left\langle I_+(\phi)^2I_-(\phi)\right\rangle_{_L}}{\left\langle I_+(\phi)\right\rangle_{_L}^3}
\end{equation}
with 
\begin{equation}
	I_{_\pm}(\phi) = \!\!\displaystyle\int_{_0}^{^L}\!\!\!\!d\varphi\,e^{\pm[V(\phi)-V(\phi\mp \varphi)]/\Theta}
\end{equation}
and 
\begin{equation}
	\left\langle g(\phi)\right\rangle_{_L}\!\!=\!\!\displaystyle\int_{_0}^{^L}\!\!\!\!d\phi\,g(\phi).
\end{equation}
In addition, the system is also characterized by its rate of entropy production, $\sigma$,
for which
$T\sigma\equiv \omega_{\rm ss}f$ can be bounded below by the thermodynamic uncertainty relation (TUR)\,\cite{BaratoSeifert2015}:
\begin{equation}
	\label{Eq:Entropy/Heirachy}
	T\varsigma_{_\textsc{tur}}=\frac{\langle\phi\rangle}{t}\left(2\Theta\frac{\langle\phi\rangle}{\langle\phi^2\rangle-\langle\phi\rangle^2}\right)\leq T\sigma =\omega_{\rm ss}f\,.
\end{equation}
The bound of the entropy rate $\varsigma_{_\textsc{tur}}$ is based on the analysis of the system's fluctuations, requiring no direct knowledge of the tilt $f$, and was initially applied in the study of Brownian motors experiencing a tilted periodic potential. From the inequality stated by Eq.\,(\ref{Eq:Entropy/Heirachy}) and the definitions in Eqs.\,(\ref{Eq:vss/Reimann})-(\ref{Eq:Dss/Reimann}), we define an effective tilt $f_0$ as 
\begin{equation}
	\label{Eq:f/TUR}
	2\Theta\frac{\langle\phi\rangle}{\langle\phi^2\rangle-\langle\phi\rangle^2} = \Theta\frac{\omega_{\rm ss}}{D_{\rm ss}} \equiv f_0 \, . 
\end{equation}
The significance of $f_0$ comes from the feature that it can be constructed directly from observable quantities and, depending on the operating parameter regime, can serve as a good approximation for the actual tilt $f$, as will be discussed below.

\section{Reconstruction of the energy landscape}

Our objective is to derive analytical expressions that allow the determination of the energy landscape, by estimating the energy barrier $E_{\rm a}$, the periodicity $L$, and the tilting torque $f$ from experimental data. 

The viscous load $\gamma$, which governs the dynamics within the potential, requires careful consideration.
Since the geometry and size of the bead tethered to the motor are known, the Stokes expression for the drag on a sphere can be applied. However, two complications arise. First, the bead's proximity to the cell surface requires hydrodynamic corrections \cite{NordPedaci2020}, which depend on the bead-surface gap, which is not experimentally accessible. Second, internal friction contributions from motor components add a hidden drag component, which has been estimated \cite{WangChenZhangEtAl2020} from a fluctuation analysis (power spectrum measurements), but remains to be verified. Crucially, two of the three methods we describe below yield $\gamma$ directly, removing the need for prior estimation.

In the following analysis, we divide the parameter space into three regimes based on the relative strength of the tilting torque, using the dimensionless parameter  
\begin{equation}
	\epsilon \equiv \frac{fL}{\pi E_{\rm a}} \, .
\end{equation}
The three regimes are defined as follows:  (i) \textit{Low-tilt} regime, $\epsilon < 1$, corresponding to near-equilibrium conditions; (ii) \textit{High-tilt} regime, $\epsilon > 1$, characterizing far-from-equilibrium dynamics; (iii) \textit{Critical-tilt} regime, $\epsilon \approx 1$, marking the crossover between these two extremes.

\subsection{Method 1}
For our first method, we adopt a formulation of the Smoluchowski equation Eq.\,(\ref{Eq:Evol/Smoluchowski/rSpace}) based on its spectral decomposition, using techniques inspired by solid-state physics. This eigenspectrum-based approach enables robust analysis and allows for the reconstruction of the unknown energy landscape from dynamic observables\,\cite{Risken1996,Gardiner2009,ChallisJack2013,Challis2016,Lopez-AlamillaJackChallis2018,Lopez-AlamillaJackChallis2020,JackLopez-AlamillaChallis2020}.
Due to the periodicity of the term $\frac{\partial}{\partial\phi}V(\phi)$ in the operator $\mathcal{L}$ in  Eq.\,(\ref{Eq:Evol/Smoluchowski/rSpace}),  the distribution $P_{\rm ss}$ is also periodic.
Eq.\,(\ref{Eq:Evol/Smoluchowski/rSpace}) can thus be expanded in a Fourier series, enabling the use of Bloch's theorem as an analytical framework\,\cite{Kittel2004, Gardiner2009}. Within this formalism, the evolution operator $\mathcal{L}$ satisfies the eigenvalue equation\,\cite{Risken1996, Gardiner2009, ChallisJack2013}
\begin{equation}
	\label{Eq:Eigen/Smoluchowski}
	\mathcal{L}\Phi_{\alpha,k}(\phi) = -\lambda_{\alpha,k}\Phi_{\alpha,k}(\phi),
\end{equation} 
where $k$ is the wave number defined within the Brillouin zone, and $\alpha\in\mathbb{N}$ is the band index. 
The eigenfunctions $\Phi_{\alpha,k}(\phi) = \exp({ik\phi})u_{\alpha,k}(\phi)$ take the Bloch form, where $u_{\alpha,k}(\phi+L) = u_{\alpha,k}(\phi)$ has the same periodicity of $V_{\rm eq}(\phi)$, and the corresponding eigenvalues $\lambda_{\alpha,k}$ are generally complex for $f \neq 0$. 
The velocity and diffusion coefficient in this formalism are determined by \cite{Festa1978}
\begin{equation}\label{Eq:vss/Eigen}
	\omega_{\rm ss} = \frac{\partial {\,\rm Im}\{\lambda_{0,k}\}}{\partial k}\bigg\rvert_{k=0}
\end{equation}
\begin{equation}\label{Eq:Dss/Eigen}
	D_{\rm ss} = \frac{1}{2}\frac{\partial^2 {\,\rm Re}\{\lambda_{0,k}\}}{\partial k^2}\bigg\rvert_{k=0}\,.
\end{equation} 
This establishes a direct connection between the observables and the underlying eigenvalue band structure. To proceed, we focus on the three different parameter regions defined by $\epsilon$ to find explicit expressions of $\omega_{\rm ss}$, $D_{\rm ss}$, and $f$.

\vspace{10pt}
\paragraph*{Low-tilt regime $(\epsilon < 1)$.}
When $\epsilon<1$ and $E_{\rm a} \gg \Theta$, the eigenstates in higher bands decay rapidly and can be neglected at long
times, corresponding to the Kramers regime. This 
approximation allows 
Eq.\,(\ref{Eq:Eigen/Smoluchowski}) to be reduced to a master equation with well localized states,  and well defined transition rates ($h_{\pm}$) to nearest-neighbors only\,\cite{ChallisJack2013,Challis2016}. 
In this regime, the steady-state speed and the rotational diffusion coefficient are given by,
$\omega_{\rm ss}^{\rm lo} = L(h_+ - h_-) $ and 
$D_{\rm ss}^{\rm lo} = L^2(h_+ + h_-)/2$ (Eqs.\,(\ref{Eq:vss/Eigen})-(\ref{Eq:Dss/Eigen})), 
where 
\begin{equation} 
	h_\pm = \left(\frac{\pi E_{\rm a}}{L^2\gamma}\right)
	e^{-E_{\rm a}/\Theta \pm fL/(2\Theta)},
\end{equation}
leading to
\begin{equation}
	\label{Eq:vss/Kramers}
	\omega_{\rm ss}^{\rm lo} = 2\sinh(fL/2\Theta)\left(\frac{\pi E_{\rm a}}{L\gamma}\right)e^{-E_{\rm a}/\Theta} \quad\quad (\epsilon < 1)
\end{equation} 
\begin{equation}
	\label{Eq:Dss/Kramers}
	D_{\rm ss}^{\rm lo}=\cosh(fL/2\Theta)\left(\frac{\pi E_{\rm a}}{\gamma}\right)e^{-E_{\rm a}/\Theta} \quad\quad (\epsilon < 1).
\end{equation} 

Using Eqs.\,(\ref{Eq:vss/Kramers}) and (\ref{Eq:Dss/Kramers}), we can further write $D_{\rm ss}$ as a function of $\omega_{\rm ss}$ as
\begin{equation}
	\label{Eq:Dss/vss/Kramers}
	D_{\rm ss}^{\rm lo}(\omega_{\rm ss}^{\rm lo}) = D_{\textsc{k}}\sqrt{1+\left(\frac{L\omega_{\rm ss}^{\rm lo}}{2D_{\textsc{k}}}\right)^2} \quad\quad (\epsilon < 1),
\end{equation}
where we have defined 
\begin{equation}
	\label{Eq:DKramers}
	D_{\textsc{k}}=(\pi E_{\rm a}/\gamma)e^{-E_{\rm a}/\Theta}, 
\end{equation}
the Kramers expression for the equilibrium diffusion coefficient, resulting from the suppression of diffusion by the potential landscape barriers. This approximation can only be justified for energy barriers that are sufficiently large compared with 
$\Theta$ (see S.I.).
As we will see, this relationship, together with the corresponding expression derived below for the case $\epsilon>1$ (Eq.\,(\ref{Eq:Dss/vss/Pertur})), will allow us to fit the experimental  
results for $D_{\rm ss}$ 
as a function of
$\omega_{\rm ss}$ to estimate the parameters $E_{\rm a}$, $L$, and $\gamma$.
To estimate the tilt torque $f$, we write the ratio $f_0 = \Theta\omega_{\rm ss}/D_{\rm ss}$ from Eq.\,(\ref{Eq:f/TUR}) \cite{Lopez-AlamillaCachi2022a} as
\begin{eqnarray}
	f_0=\Theta \frac{\omega_{\rm ss}^{\rm lo}}{D_{\rm ss}^{\rm lo}} 
	& =  & \frac{2\Theta}{L}\tanh\left(\frac{fL}{2\Theta}\right). 
	\label{Eq:Kramerexp}
\end{eqnarray}
The tilt torque $f$ can thus be expressed as
\begin{equation}
	\label{Eq:f/Kramers-1}
	f = \frac{2\Theta}{L}\tanh^{-1}
	\left(
	\frac{Lf_0}{2\Theta}
	\right),
\end{equation}
whose series expansion is given by 
\begin{equation}
	\label{Eq:f/Kramerexp-2}
	f \approx f_0
	\left[
	1 + \frac{1}{12} 
	\left(\frac{L f_0}{\Theta}\right)^2
	\right] 
	\equiv f^{\rm lo}(f_0),
\end{equation}
demonstrating that $f_0 = \Theta\omega_{\rm ss}/D_{\rm ss}$ is a good approximation for $f$ in this regime, 
as long as the first correction in Eq.\,(\ref{Eq:f/Kramerexp-2}) is small, which is true for $\epsilon \ll 2\Theta/(\pi E_{\rm a}) < 1$.  

\vspace{10pt}
\paragraph*{High-tilt regime $(\epsilon>1)$.}
In the regime $\epsilon>1$, we will employ perturbation theory to the Smoluchowski equation to obtain the corresponding expressions of $\omega_{\rm ss}, D_{\rm ss}$, and $f$. 
We consider the evolution operator Eq.\,(\ref{Eq:evol_operat}) as 
$\mathcal{L} = \mathcal{L}_{0}+\mathcal{L}_{1}$, 
where
\begin{equation}
	\gamma\mathcal{L}_{0}=\left\{ -f\frac{\partial}{\partial \phi}+\Theta\frac{\partial^{2}}{\partial \phi^{2}}\right\}
\end{equation}
and 
\begin{equation}
	\gamma\mathcal{L}_{1}=\frac{\partial}{\partial\phi} \left[\frac{\partial V_{\rm eq}(\phi)}{\partial \phi}\right].
\end{equation}
Due to the single-mode periodicity, we expand $V_{\rm eq}(\phi)$ and $u_{\alpha,k}(\phi)|_{\alpha=0}$ in Fourier series\,\cite{Lopez-AlamillaJackChallis2020}, obtaining
\begin{equation}
	\label{Matrix:Eigen}
	\bar{k}\left\{\big[\bar{k}+\!i\bar{f}\big]\hat{u}^{0,\bar{k}}_n+\frac{\bar{E_{\rm a}}}{4}\big[\hat{u}^{0,\bar{k}}_{n+1}-\hat{u}^{0,\bar{k}}_{n-1}\big]\right\} = \!\bar{\lambda}_{\alpha,\bar{k}}\hat{u}^{0,\bar{k}}_n\,,
\end{equation}
where $\hat{u}^{0,k}_n$ is the $n$-th Fourier coefficient of $u_{0,k}(\phi)$, and where we define the dimensionless quantities $\bar{k}=kL/2\pi$, $\bar{f}=fL/2\pi\Theta$, $\bar{E_{\rm a}}=E_{\rm a}/\Theta$, and $\bar{\lambda}=\gamma \lambda L^{2}/\Theta(2\pi)^2$. 
Noting that $\bar{E_{\rm a}}/\bar{f}=2/\epsilon$, the condition $\epsilon>1$ implies that $\bar{E_{\rm a}}/\bar{f}\ll 1$, and therefore perturbation theory is applicable.
Via non-degenerate perturbation theory up to second order in $\bar{E}_a$, we find that the lowest eigenvalue is
\begin{equation}
	\bar{\lambda}_{0,\bar{k}} \approx \bar{k}^2+i\bar{f}\bar{k}+\frac{\bar{E}_a}{16}^{\!2}\!\left[\frac{\bar{k}(\bar{k}+1)}{1+i\bar{f}+2\bar{k}}+\frac{\bar{k}(\bar{k}-1)}{1-i\bar{f}-2\bar{k}}\right].\label{Eq:lambda/Approx}
\end{equation}
By inserting Eq.\,(\ref{Eq:lambda/Approx}) in Eqs.\,(\ref{Eq:vss/Eigen})-(\ref{Eq:Dss/Eigen}), we can derive the explicit approximate expressions\,\cite{Lopez-AlamillaJackChallis2020}  
\begin{align}
	\label{Eq:vss/Pertur}
	\omega_{\rm ss}^{\rm hi} \approx \frac{f}{\gamma}-\frac{f}{2\gamma}\left(\frac{\pi E_{\rm a}}{2fL}\right)^2 \quad\quad (\epsilon > 1)
\end{align}
\begin{align}
	\label{Eq:Dss/Pertur}
	D_{\rm ss}^{\rm hi} \approx \frac{\Theta}{\gamma}+\frac{3\Theta}{2\gamma}\left(\frac{\pi E_{\rm a}}{2fL}\right)^{2} \quad\quad (\epsilon > 1).
\end{align}
As done in Eq.\,(\ref{Eq:Dss/vss/Kramers}) for the case $\epsilon<1$, from Eq.\,(\ref{Eq:vss/Pertur})-(\ref{Eq:Dss/Pertur}) we can write $D_{\rm ss}$ explicitly as a function of $\omega_{\rm ss}$,
\begin{equation}
	\label{Eq:Dss/vss/Pertur}
	\frac{
		{D_{\rm ss}^{\rm hi}}
		(\omega_{\rm ss}^{\rm hi})}
	{D_{\textsc{e}}}=1\!+\!6\!\left(\frac{L\omega_{\rm ss}^{\rm hi}}{2\beta}+\sqrt{1\!+\!8\left(\!\frac{L\omega_{\rm ss}^{\rm hi}}{2\beta}\!\right)^2} \right)^{-2}\, (\epsilon>1)
\end{equation}
where $D_{\textsc{e}}=\Theta/\gamma$ is the barrier-free Einstein diffusion coefficient, and we have defined $\beta=(\pi E_{\rm a}/\gamma)$ [noting that $D_{\textsc{k}}=\beta \exp(-E_{\rm a}/\Theta)$]. 
As mentioned above, the two expressions for $D_{\rm ss}(\omega_{\rm ss})$ in Eq.\,(\ref{Eq:Dss/vss/Kramers}) and Eq.\,(\ref{Eq:Dss/vss/Pertur}), for the case $\epsilon<1$ and $\epsilon>1$ respectively, will allow us to fit the experimental results for  $D_{\rm ss}(\omega_{\rm ss})$ to estimate the parameters $E_{\rm a}$, $L$, and $\gamma$.
In order to evaluate the tilt $f$ from the observables, without assuming a value for $\gamma$, we use Eqs.\,(\ref{Eq:vss/Pertur})-(\ref{Eq:Dss/Pertur}) to write
\begin{equation}
	f_0 = \Theta\frac{\omega_{\rm ss}^{\rm hi}}{D_{\rm ss}^{\rm hi}}\approx f\left[1-\frac{1}{8}\left(\frac{2\pi E_{\rm a}}{f L}\right)^2\right]+\mathcal{O}^4.
\end{equation}

Ignoring the higher-order terms, we can rearrange this expression to write $f$ as 
an explicit function of $f_0$:
\begin{equation}
	\label{Eq:f/Pertur-2}
	f\approx f_0\left[\frac{1}{2}+
	\frac{1}{2}\sqrt{1+2\left(\frac{\pi E_{\rm a}}{f_0L}\right)^2}\;\right]\equiv f^{\rm hi}(f_0).
\end{equation}

Since in this regime $\pi E_{\rm a}/(f_0 L) \approx \pi E_{\rm a}/(f L) =  1/\epsilon \ll 1$, the dominant term in  Eq.\,(\ref{Eq:f/Pertur-2}) is $f_0 = \Theta \omega_{\rm ss}/D_{\rm ss}$. We thus see that the value of the tilt torque $f$ can be approximated by  $f_0$ for both
$\epsilon\ll1$
(Eq.\,(\ref{Eq:f/Kramerexp-2})) 
and 
$\epsilon\gg1$
(Eq.\,(\ref{Eq:f/Pertur-2})).
In both these limits 
$\omega_{\rm ss}$ is proportional to $f$, and the coefficient of proportionality is given by the corresponding Einstein relation: 
$D_{\textsc{e}}/\Theta$ for $\epsilon\gg1$ (free diffusion regime) and
$D_{\textsc{k}}/\Theta$ for $\epsilon\ll 1$ (linear response regime where the fluctuation-dissipation theorem holds).

\vspace{10pt}
\paragraph*{Critical-tilt regime $(\epsilon \simeq 1)$.}
Unlike for the previous two regimes,
in the intermediate regime $\epsilon \simeq 1$, it is not easy to obtain simple separate expressions for 
$\omega_{\rm ss}$ and $D_{\rm ss}$.
Ideally, for a set of $D_{\rm ss}(\omega_{\rm ss})$ experimental results with enough values in the $\epsilon<1$ and $\epsilon>1$ limits, one can estimate $E_{\rm a}$, $\gamma$, $L$
by fitting the high and low $\epsilon$ data separately using 
Eqs.\,(\ref{Eq:Dss/vss/Kramers})
and
(\ref{Eq:Dss/vss/Pertur}). 
This would then enable us to numerically evaluate Eqs.\,(\ref{Eq:vss/Reimann})-(\ref{Eq:Dss/Reimann}) in the intermediate regime, $\epsilon\simeq1$. However, accurate evaluation for very large barriers ($E_{\rm a}>15\Theta$) can be challenging. If such is the case, 
or in the absence of sufficient experimental data in the $\epsilon<1$ and/or $\epsilon>1$ limits, interpolation techniques that connect the low-tilt and high-tilt regimes in a continuous way can be used. 
For the case of diffusion, we note that 
$D_{\rm ss}^{\rm lo}$ increases as $\omega_{\rm ss}$ increases,  $D_{\rm ss}^{\rm hi}$ increases as 
$\omega_{\rm ss}$ decreases, and the two limiting forms cross near the critical tilt $(\epsilon \simeq 1)$. In the Kramers regime (i.e., large enough barrier height, $E_{\rm a}$) for which $D_{\textsc{k}} \ll D_{\textsc{e}}$, we can therefore smoothly interpolate between low and high 
$\omega_{\rm ss}$ by treating the two limiting expressions for the diffusion coefficients, $D_{\rm ss}^{\rm lo}$ and $D_{\rm ss}^{\rm hi}$, as two electrical resistors in parallel: 
\begin{equation}
	\label{Eq:1/Dcr}
	\frac{1}{D_{\rm ss}}\approx 
	\frac{1}{D_{\rm ss}^{\rm lo}} + 
	\frac{1}{D_{\rm ss}^{\rm hi}}
	\equiv \frac{1}{D_{\rm ss}^{\rm cr}}.
\end{equation}
This interpolation leads to an easily implemented analytical expression that is in good agreement with the exact (numerical) results (see Fig.\,\ref{Fig:2}a). 

The intermediate linear behavior observed in Fig.\,\ref{Fig:2}a, where $D_{\rm ss} \approx L \omega_{\rm ss}/2$, is obtained in the strong-asymmetric, high-tilt,  limit of the Kramers regime, reached  when
$[L \omega_{\rm ss}/(2 D_{\textsc{k}})]^2 \gg 1$. The physical interpretation in terms of nearest-neighbor Kramers transition rates is simple: when the tilt is high enough, backward transitions are negligible ($h_{-} \ll h_+$), yielding $\omega_{\rm ss} \approx L h_+ $ and $D_{\rm ss} \approx L^2 h_+/2 \approx L \omega_{\rm ss}/2$. This linear behavior leads directly to an estimate of the motor periodicity, $L$, because no other system parameters are involved.

In the case of the torque, we can interpolate  between the low-tilt 
Eq.\,(\ref{Eq:f/Kramerexp-2}) 
and high-tilt 
Eq.\,(\ref{Eq:f/Pertur-2}) limits using the following Padé approximant:
\begin{equation}
	\label{Eq:f/Padde}
	\Theta\frac{\omega_{\rm ss}}{D_{\rm ss}} \approx f \frac{\left(\!\frac{\Theta}{\pi E_{\rm a}}\!\right)^2 +\frac{2}{3}\!\left(\!\frac{fL}{2\pi E_{\rm a}}\!\right)^4}{\left(\!\tfrac{\Theta}{\pi E_{\rm a}}\!\right)^2\!\!+\frac{1}{6}\!\left(\!\frac{fL}{2\pi E_{\rm a}}\!\right)^2\!\!+\frac{2}{3}\!\left(\!\frac{fL}{2\pi E_{\rm a}}\!\right)^4}\,.
\end{equation}
This expression shows that in the  
$\epsilon\simeq1$ regime the effect of the energy barriers cannot be neglected and is outside the scope of the Kramers approximation. 
In this case, the roots of Eq.\,(\ref{Eq:f/Padde}) can be used to find $f$, if 
$\Theta\omega_{\rm ss}/D_{\rm ss}$ and $E_{\rm a}$ are known (see S.I.). 
We will call such a solution $f^{\rm cr}$.

\vspace{10pt}
Finally, in the total absence of a tilt torque ($f=0$), we can estimate the typical time $\tau_L$ to diffuse one step 
$L$ over the energy barrier $E_{\rm a}$:
\begin{equation}
	\label{Eq:timeAB}
	\tau_L = 
	\frac{L^2}{D_{\textsc{k}}} =\frac{L^2\gamma}{\pi E_{\rm a}}e^{E_{\rm a}/\Theta}\,,
\end{equation}
In Fig.\,\ref{Fig:2}a and Fig.\,\ref{Fig:2}b, we show the behavior of the theoretical expressions found above for the diffusion coefficient,  $D_{\rm ss}$, as a function of the speed, $\omega_{\rm ss}$, and for the approximate tilt torque $f$, respectively compared with their true values, spanning the regions from $\epsilon<1$ (low speed) to $\epsilon>1$ (high speed). 
Fig.\,\ref{Fig:2}b shows the 
effective tilt,  $f_0$, normalized by $f$, as a function of the parameter $\epsilon$, quantifying the accuracy of $f_0$ as an approximation for the true tilt $f$ in the limiting cases of low-tilt ($\epsilon\ll1$) and high-tilt ($\epsilon\gg1$), respectively.
This figure (\ref{Fig:2}b) highlights the breakdown of the linear approximation in the intermediate regime $\epsilon \simeq 1$.

\vspace{5pt}
\paragraph*{Experimental data fitting.}
In Fig.\,\ref{Fig:2}c, we plot the experimentally measured values of $D_{\rm ss}$ and corresponding $\omega_{\rm ss}$ for each motor in our dataset, together with the best (nonlinear) fit obtained 
by fitting the approximate interpolation formula, 
$D_{\rm ss}^{\rm cr}$ 
(Eq.\,(\ref{Eq:1/Dcr})), to the data (with the periodicity $L$, friction coefficient $\gamma$, and barrier height $E_{\rm a}$ as fitting parameters). 
The best fit parameters are 
$L=0.218\pm0.014$\,rad, 
$\gamma =(0.087\pm0.005)$\,\pDrg, 
and 
$E_{\rm a}/\Theta=3.8 \pm 0.4$
(the use of the Kramers approximation in $D_{\rm ss}^{\rm cr}$,
Eq.\,(\ref{Eq:1/Dcr}), leads to an approximately 5\% overestimate of $E_{\rm a}$ with respect to the value obtained using the exact expression, Eq. SI.2, see Fig. S3(a)).
The best fit is plotted using a color code for 
$\epsilon$ that illustrates clearly the low 
$\omega_{\rm ss}$ plateau at $D_{\textsc{k}}$ for 
$\epsilon < 0.1$, the maximum near the critical tilt at $\epsilon=1$, and the high $\omega_{\rm ss}$ plateau at $D_{\textsc{e}}$ for 
$\epsilon > 10$.
The fitted periodicity of the system, $L$, 
leads to 28.8$\pm$1.8 steps per turn, in semi-quantitative agreement with the already discussed and expected 26-fold periodicity of the LP-ring. 
More data for both
slow motors ($\omega_{\rm ss} < 10$ rad/s) and
fast motors ($\omega_{\rm ss} > 1000$ rad/s)
would allow us to refine our fits. 

In Fig.\,\ref{Fig:2}d, we show the experimental estimates of the tilt $f$ as a function of the measured speed, obtained from the measured values of $\omega_{\rm ss}$ and $D_{\rm ss}$ using Eq.\,(\ref{Eq:f/Kramerexp-2}) (valid for $\epsilon<1 $) and Eq.\,(\ref{Eq:f/Pertur-2}) (valid for $ \epsilon>1$).  A linear fit of the form $f = \gamma \omega_{\rm ss}$ applied to the data yields an effective drag coefficient $\gamma = 0.12$\,\pDrg, in reasonable agreement with the value obtained independently in Fig.\,\ref{Fig:2}c given the difference in methods used. It is worth noting that the determined viscous drag is approximately an order of magnitude larger than the expected value based on the hydrodynamic drag 
$\gamma_{\textsc{b}}$ of a rotating 100 nm sphere, even after accounting for proximity corrections near a surface\,\cite{NordBiquet-BisquertAbkarianPigaglioEtAl2022}. This discrepancy likely reflects the contribution of internal drag,
$\gamma_{\textsc{R}}$,
within the motor, which has been largely neglected until now. Internal drag arises from interactions between the rotor and static structural elements of the motor, such as the LP ring. Under the low external load conditions of our experiments, we conclude that internal friction is the dominant contributor to the total viscous drag.

\subsection{Method 2}
In our second method, similar to the approach used to obtain Eq.\,(\ref{Matrix:Eigen}), we expand the Smoluchowski Eq.\,(\ref{Eq:Evol/Smoluchowski/rSpace}) as a Fourier series in the case of a general equilibrium potential with an arbitrary number of periodic modes\,\cite{Lopez-AlamillaJackChallis2018,Lopez-AlamillaJackChallis2019a,Lopez-AlamillaJackChallis2019}. 
This method of reconstruction is robust and has been widely tested for different scenarios, including high dimensional systems, outside the Kramers regime, and systems with hidden degrees of freedom\,\cite{Lopez-AlamillaJackChallis2019a,Lopez-AlamillaJackChallis2019}. In this way we obtain a set of matrix equations 
\begin{equation}
	\label{Eq:Evol/Smoluchowski/kSpace}
	-\Bigg(\Theta k-\frac{ifL}{2\pi}\Bigg)\hat{P}_k^{\rm ss}=\sum_{q}q\hat{V}_q^{\rm eq}\hat{P}_{k-q}^{\rm ss}\quad\quad k\neq0\,,
\end{equation}
\begin{equation}
	\label{Eq:vss/kSpace}
	\omega_{\rm ss}=\frac{f}{\gamma}-\frac{i2\pi}{\gamma}\sum_{q} q\hat{V}_q^{\rm eq}\hat{P}_{-q}^{\rm ss}\,,
\end{equation}
where $k$ and $q$ are the wave numbers, and $\hat{\cdot}$ denotes the Fourier coefficients. 
These equations can be used to reconstruct $V_{\rm eq}$ by rearranging Eq.\,(\ref{Eq:Evol/Smoluchowski/kSpace}) as
\begin{equation}
	\label{Eq:Veql/kSpace}
	\hat{V}_q^{\rm eq} = -\Bigg[\sum_{q}q\hat{P}^{\rm ss}_{k-q}\Bigg]^{-1}\Bigg(\Theta k-\frac{ifL}{2\pi}\Bigg)\hat{P}_k^{\rm ss} \quad\quad k\neq0 \,,
\end{equation}   
an expression that requires the experimental estimate of $P_{\rm ss}(\phi)$ and $f$. 
As discussed in Sec.\,\ref{sec:SingleMotorMeasuremets}, $P_{\rm ss}(\phi)$ can be obtained for every experimental single-motor trace (see Fig.\,\ref{Fig:1}f). 
To estimate the tilt $f$, we first make the simplifying assumption that each trace falls within either the $\epsilon\ll1$ or  $\epsilon\gg1$ regime and then approximate $f$ by $f_0$ as defined in Eq.\,(\ref{Eq:f/TUR}).
Interestingly, this method allows us to estimate the drag $\gamma$ from the experimental observables, instead of assuming a value for it. Once $V_{\rm eq}$ is reconstructed, Eq.\,(\ref{Eq:vss/kSpace}) can be used to estimate $\gamma$ from 
\begin{equation}
	\label{Eq:gamma/kSpace}
	\gamma \approx \frac{f_0}{\omega_{\rm ss}}-\frac{i2\pi}{\omega_{\rm ss}}\sum_{q} q\hat{V}_q^{\rm eq}\hat{P}_{-q}^{\rm ss}\equiv\gamma_0\,.
\end{equation}

\vspace{5pt}
\paragraph*{Experimental data fitting.}
In Fig.\,\ref{Fig:2}e (left panel) we plot the resulting $V_{\rm eq}(\phi)$ for every motor measured, together with the mean of all the reconstructed potentials. 
The resulting energy barriers are distributed from 1 to $3\Theta$, with an average of $2\Theta$, a value smaller than the 
$(3.80 \pm 0.40)\Theta$ 
found in Method 1 (but in agreement with the result of Method 3 below). The possible reasons for this 
discrepancy
are: 
i) Method 1 assumes that the system is in the Kramers regime making barriers by default larger than 3$\Theta$ \cite{Challis2016}. 
ii) The short traces may limit the accuracy of the $P_{\rm ss}$ estimation required by Methods 2 and 3\,\cite{Lopez-AlamillaJackChallis2019a}. 
iii) Method 1 naturally gives greater weight to the left and right tails of the experimental $D_{\rm ss}$ vs. $\omega_{\rm ss}$ curve, which are directly governed by  $E_{\rm a}$, unlike Methods 2 and 3, which rely on a simple arithmetic mean.
In Fig.\,\ref{Fig:2}f we show the experimental distribution of the values of $\gamma$, calculated for each motor. The mean value of the distribution is $\gamma=0.13$\,\pDrg, consistent with the estimates of Method 1, highlighting again the dominant role of the motor's internal drag, 
$\gamma_{\textsc{R}}$,
over the hydrodynamical drag of the attached bead,
$\gamma_{\textsc{B}}$,
for these experiments (see Fig.\,\ref{Fig:1}d).

\subsection{Method 3}
\label{sec:method3}
Originally developed as a variant of method 2 in the 
absence of information for $f$\,\cite{Lopez-AlamillaJackChallis2019a}, this method involves an iterative self-consistent solution of Eqs.\,(\ref{Eq:vss/kSpace})-(\ref{Eq:Veql/kSpace}), and requires as inputs the experimental values for the angular distribution $P_{\rm ss}(\phi)$, angular velocity $\omega_{\rm ss}$, and $\gamma$. The optimization algorithm works as follows:
\begin{enumerate}
	\item We obtain $P_{\rm ss}$, $\omega_{\rm ss}$, $\gamma$.
	\item $f^{(i)}\equiv \gamma\omega_{\rm ss}$ serves as initial guess.
	\item Eq.\,(\ref{Eq:Veql/kSpace}) with $P_{\rm ss}$, $f^{(i)}$ returns $V_{\rm eq}^{(i)}$.
	\item Eq.\,(\ref{Eq:vss/kSpace}) with $V_{\rm eq}^{(i)}$, $f^{(i)}$, $\gamma$ returns $\omega^{(i)}$. 
	\item We check $|\omega_{\rm ss}-\omega^{(i)}|\leq \delta$, with $0\leq\delta\ll1$ a set tolerance.
	\begin{enumerate}
		\item True: $f^{(i)}\equiv f_{\rm opt}$, $V_{\rm eq}^{(i)}\equiv V_{\rm eq}^{\rm opt}$ 
		\item False: we update the input $f^{(i+1)}=f^{(i)}+\Delta f$, iterating steps 3-5 until $|\omega_{\rm ss}-\omega^{(i+1)}|\leq \delta$.
	\end{enumerate}
\end{enumerate}
Because this method is not limited to a specific $\epsilon$ regime, it therefore allows us to analyze all measured motors. Based on the results from Methods 1 and 2 above, we take $\gamma \approx 0.12$\,\pDrg\ to account for
both the internal motor drag and the external load of the attached bead. 

\begin{figure*}
	\centering
	\includegraphics[width=0.99\textwidth]{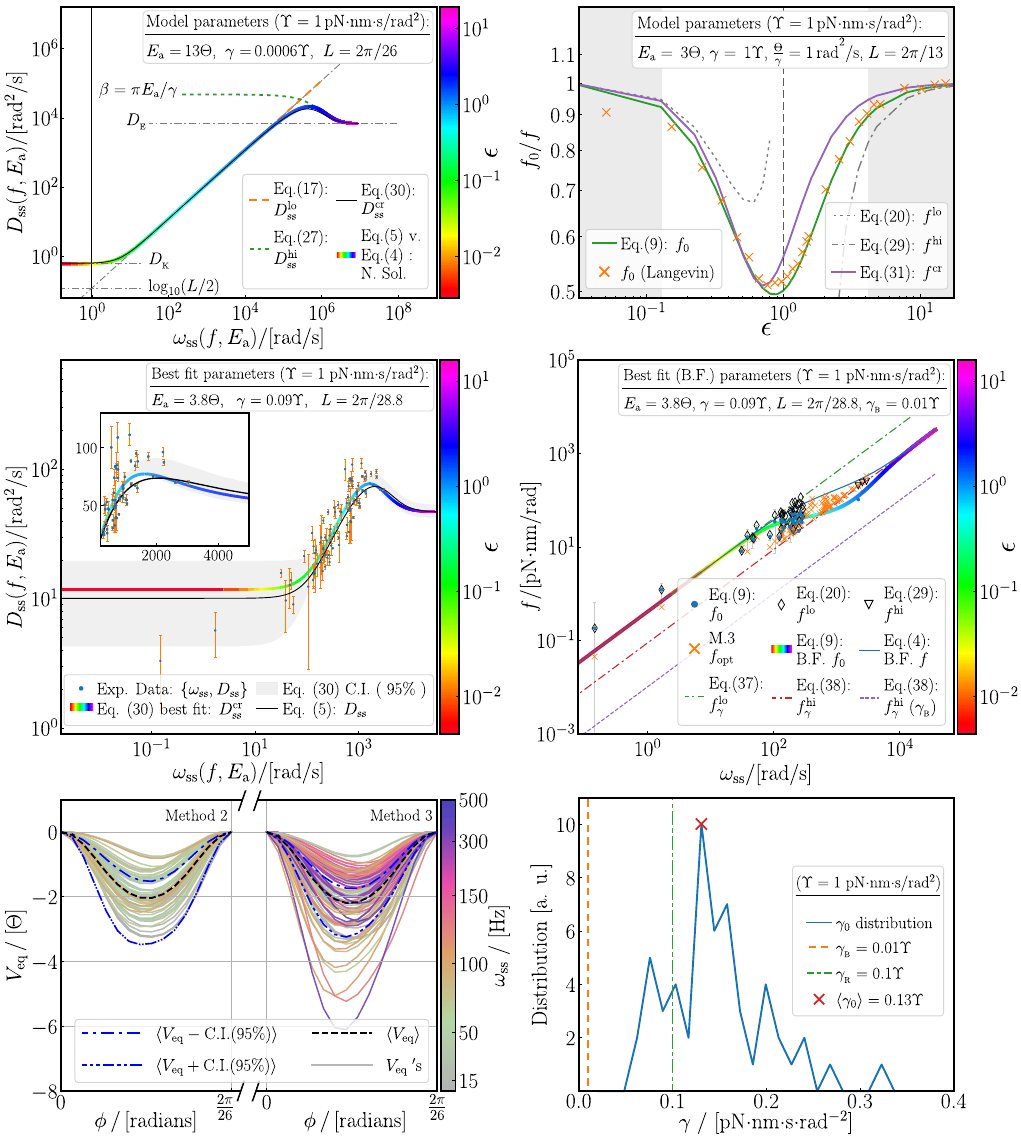}
	\put(-472,548){(a)}
	\put(-215,548){(b)}
	\put(-472,373){(c)}
	\put(-215,373){(d)}
	\put(-472,157){(e)}
	\put(-215,157){(f)}
	\put(-210,480){$\frac{f_0}{f}$}
	\put(-198,482){$_{>0.9}$}
	\put(-64,480){$\frac{f_0}{f}$}
	\put(-53,482){$_{>0.9}$}
	\caption{
		Illustrative theoretical results for Method 1 and application to experimental data to reconstruct: $E_{\rm a}$ (methods 1 to 3), $\gamma$ (methods 1, 2), effective torque (methods 1, 3), and $V_{\rm eq}$ (methods 2, 3). 
		(a) Theoretical parametric plot $D_{\rm ss}$ vs. $\omega_{\rm ss}$ along with the method 1 approximations, Eqs.\,(\ref{Eq:Dss/vss/Kramers}),\,(\ref{Eq:Dss/vss/Pertur}), (\ref{Eq:1/Dcr}), used to determine
		$D_{\rm ss}$ for a model potential. 
		(b)  Relative error between $f$ and the Method 1 torque estimates, Eqs.\,(\ref{Eq:f/Kramerexp-2}),\,(\ref{Eq:f/Pertur-2}), (\ref{Eq:f/Padde}), as a function of $\epsilon$ for a model potential.
		(c) Parametric plot $D_{\rm ss}$ vs. $\omega_{\rm ss}$ for the experimental data and best fit  model (rainbow curve parametrized by $\epsilon$) using the interpolation    Eq.\,(\ref{Eq:1/Dcr}). The shaded gray area shows the 95\% confidence interval; the black line is the exact numerical result for the same parameters used in the fit. (d) Corresponding torque values obtained from the experimental 
		$\omega_{\rm ss}$ values ($f_0$, corrected models, $f_{\rm opt}$, Eqs.\,(4)\,(9) best fit and asymptotic behavior included). (e) Resulting energy landscapes from individual traces using (left) method-2 and (right) method-3, color bar represents the corresponding experimental $\omega_{\rm ss}$ values. 
		(f) Distribution of $\gamma_0$ values calculated from $f_0$ and method-2 barriers. 
		In all panels, the model parameters and best fit parameters are listed in their respective legends. 
	}
	
	\label{Fig:2}
\end{figure*}
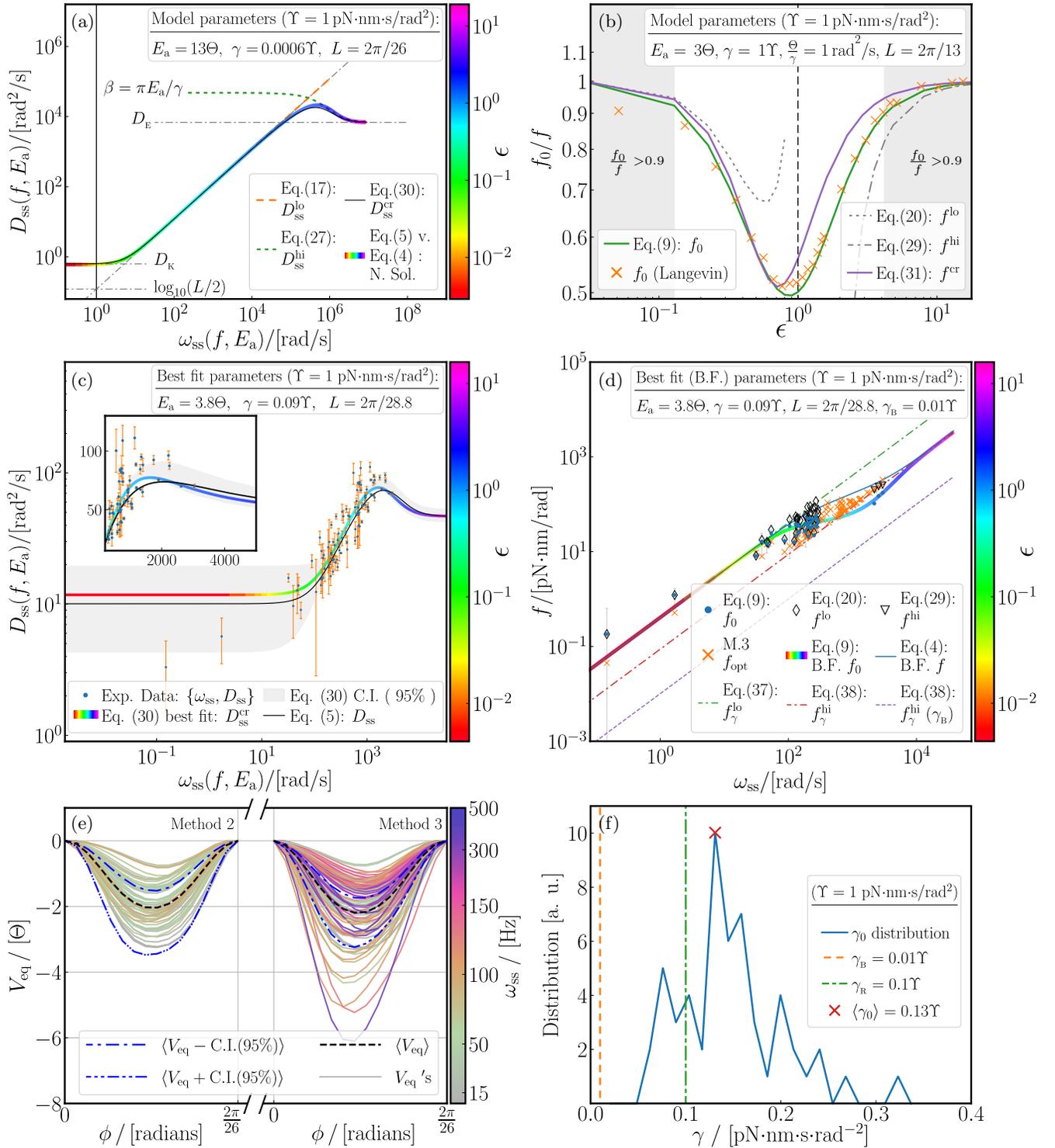

\vspace{5pt}
\paragraph*{Experimental data fitting.}
In Fig.\,\ref{Fig:2}e (right panel) we plot the results of the reconstruction of the potential $V_{\rm eq}$, obtained for each motor. 
The energy barriers and their average ($E_{\rm a}\sim 2.0\pm1.2\, \Theta$), despite a larger spread, are in good agreement with the results of Method 2 
(Fig.\,\ref{Fig:2}e, left panel). Since Methods 2 and 3 share the same procedure to estimate $P_{\rm ss}$, it is likely that this common step is the main source of the comparatively low energy barriers obtained relative to Method 1.

\section{Discussion}

Understanding how molecular motors transduce (electro-)chemical energy into directed mechanical motion requires not only identifying the molecular components involved and their interactions, but also characterizing the effective energy landscape in which they operate, determined by periodic structural constraints, and internal mechanical resistance at the atomic scale. In this work we developed and applied three complementary methods of analysis to quantitatively characterize the energy landscape that shapes the rotational dynamics of molecular motors operating in the presence of
periodic potentials. We applied these methods to the BFM, where the LP ring acts as a molecular bearing imposing a 26-fold periodic potential landscape on the rotor, which dominates the BFM's dynamics at low loads. Crucially, our approach requires neither the application of external torque, nor prior knowledge of the system’s viscous drag or the motor’s torque output.

The three methods developed in this work, along with the results they provide when applied to BFM rotation data, are summarized in Table 1. Method 1 
assumes a single Fourier mode (cosine potential) and
uses analytical solutions to the Smoluchowski equation in the limits of weak and strong tilt Eqs.\,(\ref{Eq:Dss/vss/Kramers}), (\ref{Eq:Dss/vss/Pertur}), (\ref{Eq:1/Dcr}) to relate the measured
rotational velocity and diffusion to the viscous drag, periodicity, and energy barrier. This method can also be used to relate the measured rotational velocity and diffusion to the
underlying torque via Eqs.\,(\ref{Eq:f/Kramers-1}), (\ref{Eq:f/Pertur-2}), (\ref{Eq:f/Padde}).
Method 2 employs a potential landscape model-independent, Fourier-based reconstruction of the Smoluchowski equation, by fitting the full steady-state angular distribution $P_{\rm ss}$, along with velocity and diffusion, to extract the underlying potential and drag. Method 3 provides a direct reconstruction of the energy landscape by integrating $P_{\rm ss}$ under the assumption of known viscous drag.

The viscous drag coefficient experienced by the motor was estimated as $\gamma\approx0.1$\,\pDrg, with consistent values obtained from Methods 1 and 2.  This result, which is in quantitative  agreement with a previous estimate \cite{WangChenZhangEtAl2020} of internal drag,  suggests that internal structural interactions, including those at the rod-LP ring interface, can contribute significantly to mechanical resistance and rotational dissipation. We conclude that, in the low-load regime, the hydrodynamic drag of the bead, often estimated using the Stokes formulation with surface proximity corrections \cite{NordBiquet-BisquertAbkarianPigaglioEtAl2022}, contributes only a small portion of the total measured resistance. The corresponding drag coefficient 
($\sim$0.008-0.01\,\pDrg) implies that the internal viscous friction, one order of magnitude higher, is the primary source of dissipation 
in the far-from-equilibrium regime. In the close-to-equilibrium
regime, on the other hand, a large enough potential landscape energy barrier plays a major role in determining  the diffusion coefficient. Therefore, in both limiting scenarios, very-low (very-high) tilts, two different linear asymptotic limits for the torque ($f\propto\omega_{\rm ss}$) can be deduced. The corresponding  proportionality constants are related to the barrier-suppressed diffusion, $D_{\textsc{k}}$, and the barrier-free diffusion, $D_{\textsc{e}}$, respectively:
\begin{equation}\label{Eq:f/gammaK}
	f_\gamma^{\rm lo}\approx\frac{\Theta}{D_{\textsc{k}}}\omega_{\rm ss}=\gamma\frac{\Theta}{\pi E_{\rm a}}e^{E_{\rm a}/\Theta}\omega_{\rm ss},
\end{equation}
\begin{equation}\label{Eq:f/gamma}
	f_\gamma^{\rm hi}\approx\frac{\Theta}{D_{\textsc{e}}}\omega_{\rm ss}=\gamma\omega_{\rm ss}.
\end{equation}
These asymptotic limits are displayed in Fig.\,\ref{Fig:2}d using the best-fit parameters found by method 1. For contrast, this plot also included the case where $f_\gamma^{\rm hi}$ only accounts for the bead's viscous drag $\gamma_{\textsc{b}}$, highlighting again the need to include the internal 
drag contribution at low loads.  

Cryo-EM structures reveal a tight interface between the rod and LP ring, involving both electrostatic and steric interactions
\cite{JohnsonFurlongDemeNordCaesarChevanceBerryHughesLea2021, Yamaguchi2021}. Opposing charges and potential salt bridges likely stabilize the rotor and support high-speed rotation, while also introducing periodic energy barriers that contribute to the internal resistance and shape motor dynamics. All three of our methods enabled estimates of the height of the LP ring-rod energy barrier. Method 1 gave a barrier height of 
$E_{\rm a} = 3.8\,\Theta$. Method 2 yielded a distribution of \( E_{\rm a} \) values between 1 and \( 3\,\Theta \), with an average around \( 2\,\Theta \), and Method 3 produced a similar average of \( E_{\rm a} \sim 2\,\Theta\).  Our estimates for
$E_{\rm a}$ and $\gamma$ are in qualitative agreement with those of a previous modeling study\,\cite{MoraYuSowaWingreen2009} in which, contrary  to our own study, a 26-fold periodicity was assumed as input to investigate the stepping of the BFM. The only other experimental measurement of this barrier height was recently reported by Rieu et al., who used high-speed back-scattering darkfield polarization microscopy to track 
with exceptional angular and temporal resolution
the equilibrium diffusion 
of gold nanorods 
(i.e., in the absence of applied torque)
attached to the bacterial hook  \cite{RieuNordCourbetElSayyedEtAl2025}. By analyzing angular hopping events, they inferred energy barriers on the order of \( 13\,\Theta \). 
We offer a few speculative explanations for why our estimates differ from
those of Rieu et al. In their study, the authors estimated \( E_{\rm a} \) by applying Kramers’ theory to measured dwell times between angular states, a method that relies on assumptions about the shape of the energy landscape. They assumed a 
periodic parabolic barrier shape, which leads to a cusp at the peak and a harmonic form at the potential minimum. This choice modifies the prefactor appearing in Kramers’ rate 
and leads to an overestimate of the barrier height with respect to the one obtained for a cosine potential (we estimate this overestimate to be less than 20\% for barrier heights between 4 and 13 $\Theta$).  Secondly, and probably principally, their calculation includes only the hydrodynamic drag of the rotating rod, neglecting internal viscous drag of the motor itself. As mentioned above, our results indicate that this internal drag dominates under low-load conditions, and omitting it would lead to an overestimate for \( E_{\rm a} \). 
In the S.I. Fig.\,\ref{Fig:Same_Dk_diff_g}, we explore how the different choices of $\gamma$ for a fixed equilibrium diffusion value lead to different energy barriers. Finally, their measurements were performed in the absence of stator units, whereas our motors were actively rotating with one or two stators engaged. It is possible, and remains to be seen, whether dynamic coupling with stator units modulates the effective potential during active rotation, thereby reducing the observed barrier height.

The energy landscape of this molecular bearing raises interesting questions about its evolutionary tuning and biological function. One may imagine that evolution has tuned the barrier height to balance structural stability, maintaining alignment of the rod during high-speed rotation, with minimal mechanical resistance, thus maximizing efficiency. Depending on the number and duty-ratio of engaged stator units, we could imagine that this landscape may also act as a ratcheting mechanism, providing resistance to backward rotation when torque output is low or when stator units transiently disengage. This may be especially important in species like \textit{E. coli}, where stator assembly is dynamic \cite{Nord2017, Leake2006, LeleHosuBerg2013, TippingDelalezLimBerryArmitage2013}, compared to motors like \textit{Campylobacter jejuni}, where stators seem stably assembled \cite{Beeby2016, Drobnic2025}. 
In systems that experience mechanical strain, such as during flagellar (un-)wrapping \cite{Cohen2024}, one could imagine that larger energy barriers may be advantageous, if they enhance the structural rigidity of the rod-bearing interface.  Motility strategy may also shape barrier height, as peritrichously-flagellated bacteria and polar-flagellated species face distinct mechanical challenges that may select for different optimal landscapes. Finally, even subtle variations in rod-LP ring alignment, post-translational modifications, or local symmetry could generate cell-to-cell variability in the energy landscape, leading to cell-to-cell diversity in motor behavior and motility. We believe that the methods introduced here could provide a potential landscape model-independent means of 
quantitatively probing some of these questions, as well as offer a general framework for probing energy transduction in other molecular motors. 

\onecolumngrid


\begin{table*}[t] 
	\centering
	\caption{Comparison of the three methods developed for extracting torque, viscous drag, and the potential landscape from bacterial flagellar motor bead assay data.}
	\renewcommand{\arraystretch}{1.4}
	\begin{tabular}{|>{\raggedright\arraybackslash}p{0.13\textwidth}|
			>{\raggedright\arraybackslash}p{0.28\textwidth}|
			>{\raggedright\arraybackslash}p{0.29\textwidth}|
			>{\raggedright\arraybackslash}p{0.265\textwidth}|}
		\hline
		\textbf{Methodology} & \textbf{Method 1:\newline Spectral decomposition of Smoluchowski
			equation via Bloch's theorem} & \textbf{Method 2:\newline Fourier expansion of the Smoluchowski equation in the case of a general equilibrium
			potential} & \textbf{Method 3:\newline Iterative self-consistent landscape inversion} \\
		\hline
		\textbf{Inputs and Relevant Equations} & $\omega_{\rm ss},\,D_{\rm ss}$; \newline 
		Eqs.\,\ref{Eq:Dss/vss/Kramers}, \ref{Eq:f/Kramerexp-2}, \ref{Eq:Dss/vss/Pertur}, \ref{Eq:f/Pertur-2} & 
		$P_{\rm ss}(\phi),\,L,\,\omega_{\rm ss},\,D_{\rm ss}$; \newline 
		Eqs.\,\ref{Eq:Veql/kSpace}, \ref{Eq:gamma/kSpace} & $P_{\rm ss}(\phi),\,L,\,\omega_{\rm ss}$,\,$\gamma$; \newline 
		Eqs.\,\ref{Eq:vss/kSpace}, \ref{Eq:Veql/kSpace} \\
		\hline
		\textbf{Outputs} & $f(\omega_{ss})$, $\gamma$, $E_{\rm a}$, $L$ &   
		$V_{\rm eq}(\phi)$, $\gamma_0$, $E_{\rm a}$, $f_0$ & $V_{\rm eq}(\phi)$, $E_{\rm a}$, $f$\\
		\hline
		\textbf{Assumptions and Notes} & Valid for $\epsilon<1$ or $\epsilon>1$; \newline 
		$f \approx f_0$; \newline
		Single-mode sinusoidal potential & 
		Valid for $\epsilon<1$ or $\epsilon>1$; \newline 
		$f \approx f_0$; \newline
		Valid for arbitrary periodic potentials & Valid for all $\epsilon$;\newline 
		$\gamma=$ 0.12\,\pDrg;\newline  Valid for arbitrary periodic potentials\\
		\hline
		\textbf{Experimental Results} & 
		$\gamma = (0.09\pm0.005)$\,\pDrg 
		\newline
		$E_{\rm a} = (3.8\pm0.4)\,\Theta$ 
		\newline 
		$2\pi/L =(28.8\pm1.8)\,\mathrm{steps}$ & $\langle\gamma_0\rangle = 
		(0.13\pm 0.07)$\,\pDrg 
		\newline
		$E_{\rm a} \approx (2\pm 1.2)\,\Theta$ & $ E_{\rm a} = (2 \pm 1.2)\,\Theta $ \\
		\hline
		\textbf{Pros} & Traces analyzed can be relatively short. Does not require knowledge of $\gamma$. & A single trace is enough for applicability. Does not require knowledge of $\gamma$. & A single trace is enough for applicability. Does not require knowledge of $f$.\\
		\hline
		\textbf{Cons} & Requires multiple traces spanning across different $\epsilon$ regimes.
		\newline 
		Imposes
		the shape of the potential.&  Accurate estimation of $P_{\rm ss}$ requires motor's trace with sustained constant $\omega_{\rm ss}$, $D_{\rm ss}$ during large recording times.\newline Not applicable in the $\epsilon\simeq1$ case.& Accurate estimation of $P_{\rm ss}$ requires motor's trace with sustained constant $\omega_{\rm ss}$, $D_{\rm ss}$ during large recording times.\newline Requires knowledge of $\gamma$.\\
		\hline
	\end{tabular}
	\label{tab:methods}
\end{table*}

\twocolumngrid

\section{Materials and methods}
\begin{figure}[htbp]
	\centering
	\includegraphics[height=175pt]{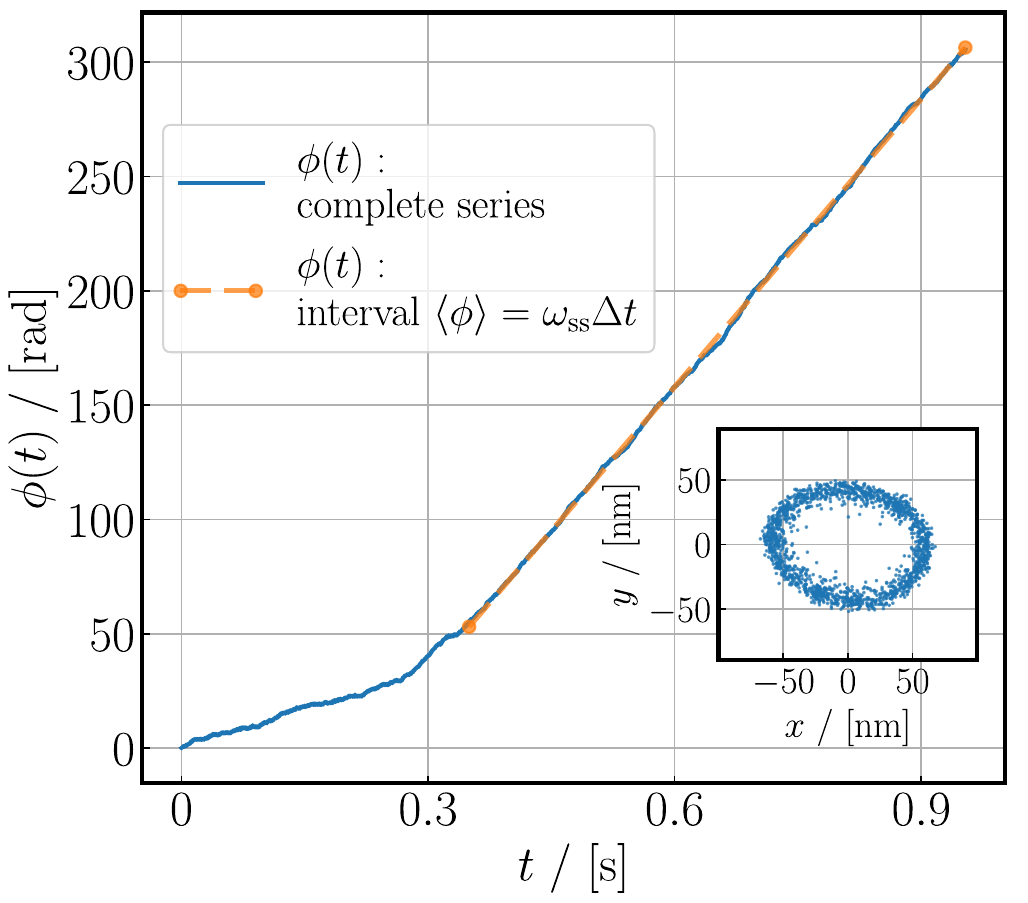}\put(-165,164){(a)}
	
	\includegraphics[height=175pt]{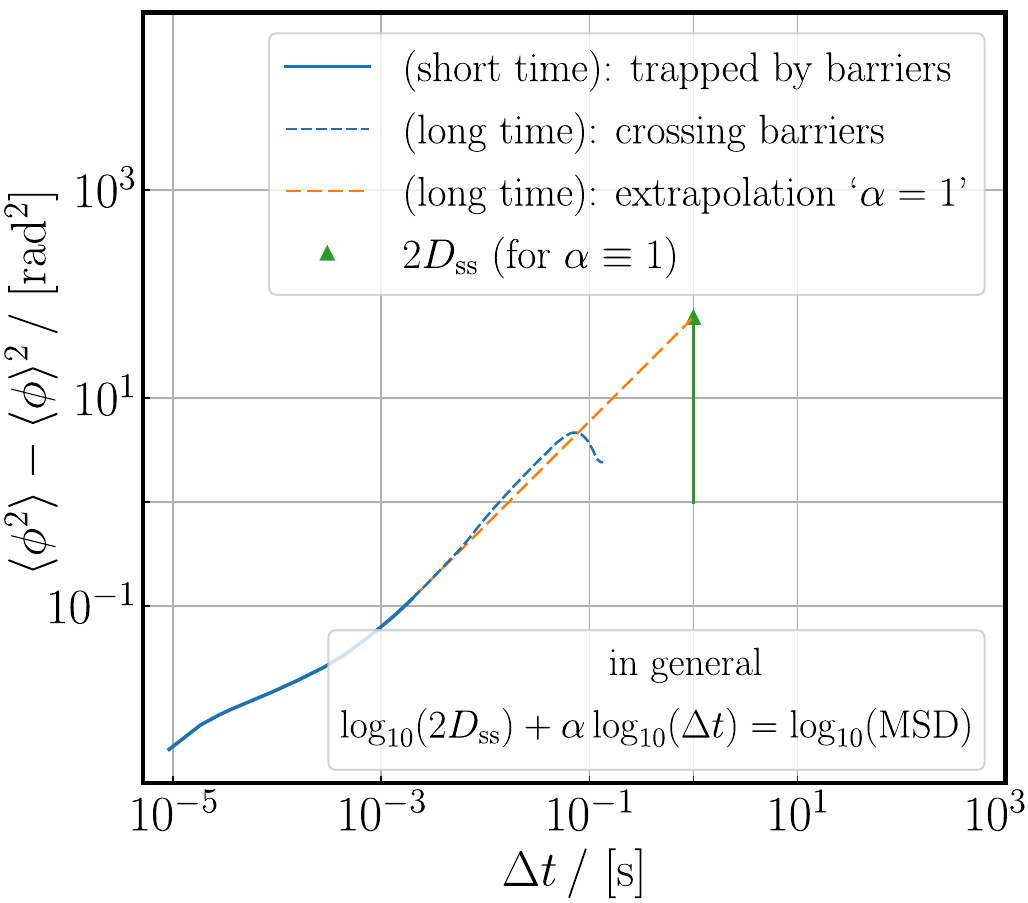}\put(-165,164){(b)}
	\caption{Example trace showing showing: (a) (solid) accumulated 
		angle versus time plot, (dashed) longest uninterrupted portion used to calculate $\omega_{\rm ss}$,  and inset $\{x(t),y(t)\}$ coordinates every 60 points, 
		(b) MSD analysis for increasing $\Delta t$ used to determine $D_{\rm ss}$.}
	\label{Fig:PreAnalysis_ss}
\end{figure}

\subparagraph{Data set.}
We measured the steady-state rotation of 95 individual BFMs of a non-switching strain of \textit{Escherichia coli} via a bead assay \cite{Hoffmann2025}, tracking the rotation of $100$ nm diameter gold beads attached to the flagellar hook, as previously described \cite{Nord2016}. The BFMs were driven by either wildtype 
\textit{E. coli} stator units or chimeric Na$^+$ stator units. Part of this dataset was reported in\,\cite{Nord2016}, while the remaining traces are newly analyzed here. The plasmon-enhanced scattered light from the gold bead was imaged using backscattering dark-field microscopy onto a high-speed camera at 109,500 frames per second.  The bead center was tracked as it rotated slightly off-axis, yielding its circular $x(t), y(t)$ trajectory (Fig. \ref{Fig:PreAnalysis_ss}a inset). Once centered at the origin, the trajectory yielded the time trace of the BFM rotation angle, $\phi(t) = \arctan(y(t)/x(t))$, from which we computed the angular speed and diffusion. 

\subparagraph{Calculation of $\bm{\omega_{\rm ss}}$, $\bm{D_{\rm ss}}$.}
For each motor, we extracted the steady-state angular speed, $\omega_{\rm ss} = \langle \phi \rangle / \Delta t$, and the rotational diffusion coefficient, $D_{\rm ss} = \left( \langle \phi^2 \rangle - \langle \phi \rangle^2 \right) / (2 \Delta t)$. Both were computed from the longest uninterrupted portion of the trace exhibiting constant speed (identified by visual inspection, typically lasting between 0.3 s and 2 s, see for example Fig. \ref{Fig:PreAnalysis_ss}a). 
To estimate the steady-state speed, we performed a linear regression $\langle \phi\rangle=\omega_{\rm ss}\Delta t + b$  for increasing $\Delta t$ for each trace over the selected time interval (cf. Fig. \ref{Fig:PreAnalysis_ss}a). For the diffusion analysis, the mean-squared angular displacement (MSD) was computed as $\left( \langle \phi^2 \rangle - \langle \phi \rangle^2 \right)=$\,MSD\,$=2D_{\rm ss}\Delta t^\alpha$, where $\alpha=1$ corresponds to normal diffusion and $\alpha\neq1$ to anomalous diffusion. 

\begin{figure}[htbp]
	\centering 
	\includegraphics[height=175pt]{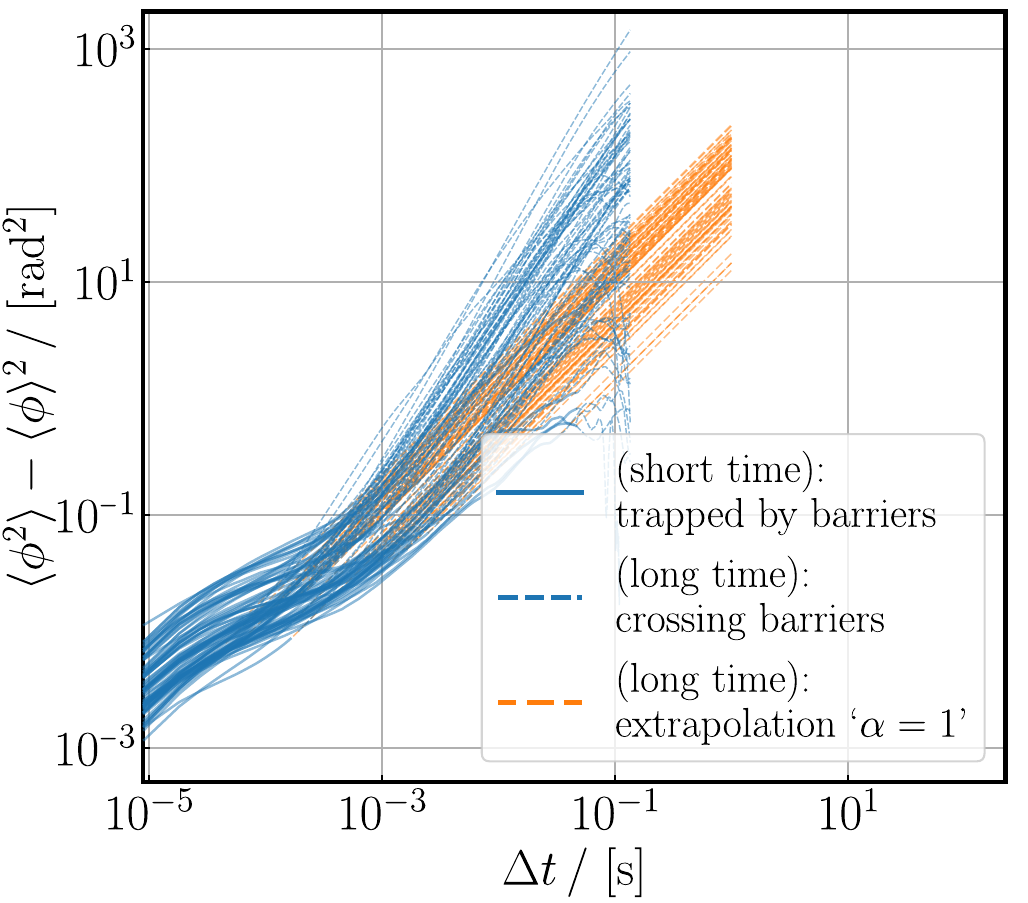}
	\caption{Set of traces showing: MSD analysis for increasing $\Delta t$ used to determine $D_{\rm ss}$.}
	\label{Fig:Analysis_ss_Ensemble}
\end{figure}

In the data analyzed here, the MSD exhibited normal diffusive behavior within certain time windows ($\Delta t \sim$1ms-0.1s). At longer time windows, deviations from stationary velocity give rise to deviations from the normal diffusion regime. Thus, $D_{\rm ss}$ was obtained by performing a linear regression within the $\Delta t \sim$1ms-0.1s windows, 
where Eqs.\,(\ref{Eq:Dss/Reimann}) and (\ref{Eq:Dss/Kramers}) hold, see Fig.\,\ref{Fig:PreAnalysis_ss}b (a detailed determination of the $\alpha=1$ region and its validity are discussed in Ref.\,\cite{LanoiseleePagniniWylomanska2025}). Figure\,\ref{Fig:PreAnalysis_ss} shows an example trace and the extraction of $\omega_{\rm ss}$ and $D_{\rm ss}$, while Fig.\,\ref{Fig:Analysis_ss_Ensemble} summarizes the $D_{\rm ss}$ calculation for the entire dataset.

\subparagraph{Calculation of the fundamental periodicity using methods 2 and 3.}
To determine the fundamental periodicity of the motor and identify the angular step size from experimental traces in methods 2 and 3, we first computed the histogram of the angular position $\phi(t)$, which we call 
$H(\phi)\big\rvert_{2\pi} 
\equiv H(\phi \, \rm{mod} \, 2\pi)$, where $\phi(t) \in [0, 2\pi]$. Because $\phi(t)$ is derived from experimental position tracking, it reflects both biological variability (e.g., fluctuations in motor dynamics) and experimental noise (e.g., imaging and tracking uncertainty), which together make it harder to identify any underlying periodicity in the histogram. To reduce this noise and extract underlying
physical periodic features, we applied Ensemble Empirical Mode Decomposition (EEMD) to the time series $\phi(t)$. EEMD is an adaptive filtering technique designed to decompose non-stationary and nonlinear signals into a sum of intrinsic mode functions (IMFs). Each IMF represents oscillatory components at different timescales, allowing us to identify dominant periodicities in the signal without requiring prior assumptions about the signal's frequency content.

\begin{figure}[htbp]
	\centering
	\includegraphics[height=270pt]{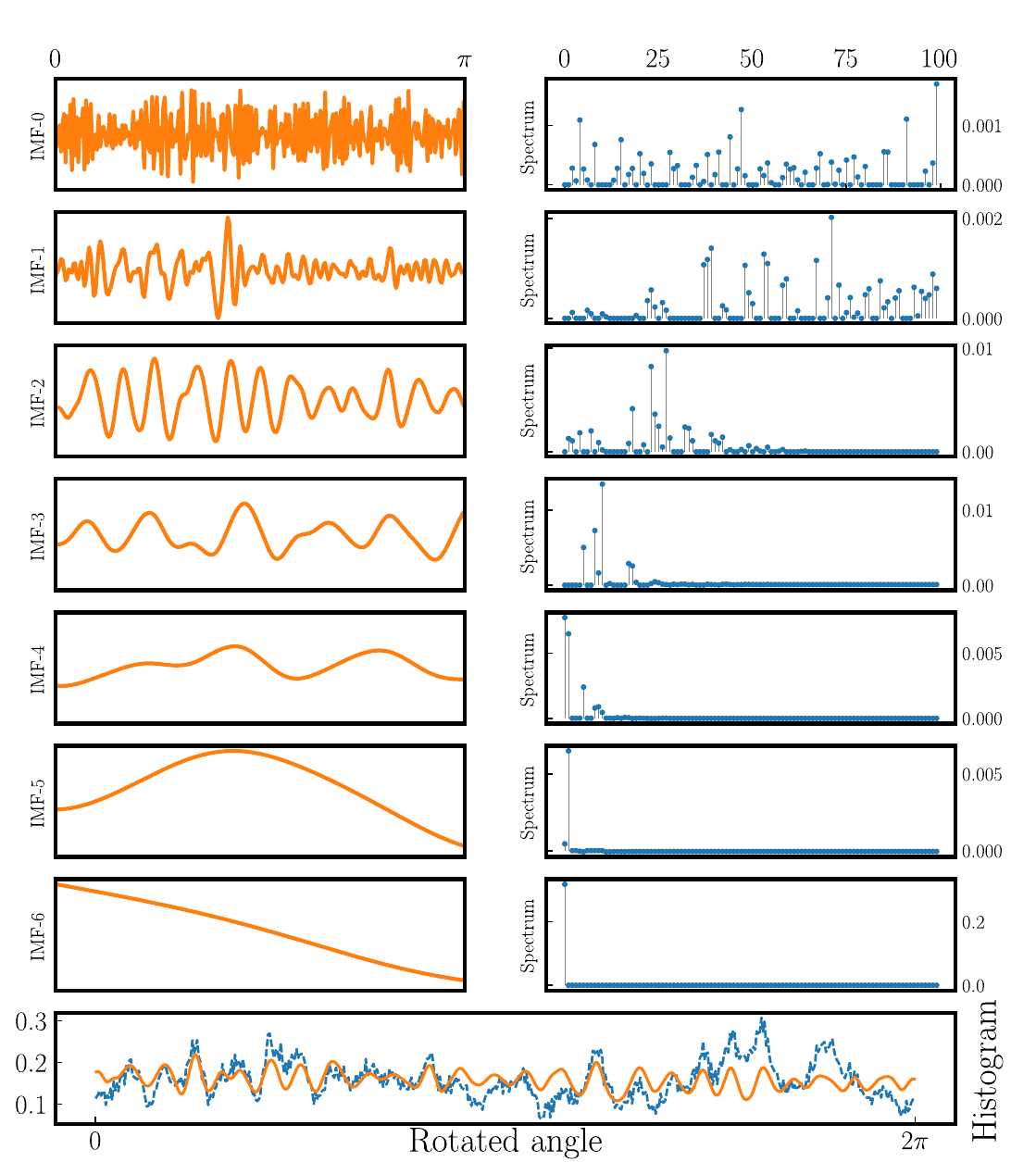}\put(-190,263){(a)}\put(-77,263){(b)}\put(-37,24){(c)}
	\caption{EEMD analysis of the trace shown in Fig. \ref{Fig:PreAnalysis_ss}a. (a) The IMFs ordered in decreasing frequency, (b) the corresponding spectra as function of the steps per rotation, and (c) angular histogram of the original signal $H(\phi)\big\rvert_{2\pi}$ (dashed line) and the filtered signal $H(\phi)^\dagger\big\rvert_{2\pi}$ (solid line).}
	\label{Fig:EMDAnalysis_ss}
\end{figure}

In Fig. \ref{Fig:EMDAnalysis_ss}, we show how the original angular histogram $H(\phi)\big\rvert_{2\pi}$ is decomposed into a set of IMFs using EEMD, ordered from high to low frequency content. Across all traces, the second mode (IMF-2) exhibits a clear 26-fold periodicity (see Fig. \ref{Fig:1}e). The sixth mode (IMF-6) captures the slow baseline variation associated with the overall shape of the original distribution (i.e., the normalization envelope). Other IMFs were excluded from further analysis. We reconstructed a denoised version of the angular histogram, $H(\phi)^\dagger\big\rvert_{2\pi}$, using only IMF-2 and IMF-6. From this, we identified the fundamental periodicity as $L = 2\pi/25.7\approx2\pi/26$, which we then used to compute an approximate steady-state angular probability distribution, $P_{\rm ss}(\phi) \approx H(\phi)^\dagger\big\rvert_{L}$, shown in Fig.\,\ref{Fig:1}f.

\stoptoc
\begin{acknowledgments}
	The authors thank Richard M. Berry, in whose lab the data were acquired. NJLA, N-OW, JP, and FP were supported by ANR BaElMec (ANR-23-CE30-0010). NJLA, N-OW, and JP were supported by the LabEx NUMEV (ANR-10-LABX-0020) within the I-Site MUSE (ANR-16-IDEX-0006). ALN was supported by the ANR PHYBABIFO (ANR-22-CE30-0034) and PHYBION (ANR-21-CO15-0004) project grants, as well as by the “Transatlantic Research Partnership", a program of the FACE Foundation and the French Embassy. The CBS is a member of the France-BioImaging (FBI) and the French Infrastructure for Integrated Structural Biology (FRISBI), two national infrastructures supported by the French National Research Agency (ANR-10-INBS-04-01 and ANR-10-INBS-05, respectively). All authors were supported by the Centre Nationale de Recherche Scientifique (CNRS) and the University of Montpellier.
	
\end{acknowledgments}


\begin{thebibliography}{48}%
	\makeatletter
	\providecommand \@ifxundefined [1]{%
		\@ifx{#1\undefined}
	}%
	\providecommand \@ifnum [1]{%
		\ifnum #1\expandafter \@firstoftwo
		\else \expandafter \@secondoftwo
		\fi
	}%
	\providecommand \@ifx [1]{%
		\ifx #1\expandafter \@firstoftwo
		\else \expandafter \@secondoftwo
		\fi
	}%
	\providecommand \natexlab [1]{#1}%
	\providecommand \enquote  [1]{``#1''}%
	\providecommand \bibnamefont  [1]{#1}%
	\providecommand \bibfnamefont [1]{#1}%
	\providecommand \citenamefont [1]{#1}%
	\providecommand \href@noop [0]{\@secondoftwo}%
	\providecommand \href [0]{\begingroup \@sanitize@url \@href}%
	\providecommand \@href[1]{\@@startlink{#1}\@@href}%
	\providecommand \@@href[1]{\endgroup#1\@@endlink}%
	\providecommand \@sanitize@url [0]{\catcode `\\12\catcode `\$12\catcode
		`\&12\catcode `\#12\catcode `\^12\catcode `\_12\catcode `\%12\relax}%
	\providecommand \@@startlink[1]{}%
	\providecommand \@@endlink[0]{}%
	\providecommand \url  [0]{\begingroup\@sanitize@url \@url }%
	\providecommand \@url [1]{\endgroup\@href {#1}{\urlprefix }}%
	\providecommand \urlprefix  [0]{URL }%
	\providecommand \Eprint [0]{\href }%
	\providecommand \doibase [0]{https://doi.org/}%
	\providecommand \selectlanguage [0]{\@gobble}%
	\providecommand \bibinfo  [0]{\@secondoftwo}%
	\providecommand \bibfield  [0]{\@secondoftwo}%
	\providecommand \translation [1]{[#1]}%
	\providecommand \BibitemOpen [0]{}%
	\providecommand \bibitemStop [0]{}%
	\providecommand \bibitemNoStop [0]{.\EOS\space}%
	\providecommand \EOS [0]{\spacefactor3000\relax}%
	\providecommand \BibitemShut  [1]{\csname bibitem#1\endcsname}%
	\let\auto@bib@innerbib\@empty
	\bibitem [{\citenamefont {Alberts}\ \emph {et\,al.}(2014)\citenamefont
		{Alberts}, \citenamefont {Johnson}, \citenamefont {Lewis}, \citenamefont
		{Raff}, \citenamefont {Roberts},\ and\ \citenamefont {Walter}}]{Alberts2014}%
	\BibitemOpen
	\bibfield  {author} {\bibinfo {author} {\bibfnamefont {B.}\,\bibnamefont
			{Alberts}}, \bibinfo {author} {\bibfnamefont {A.}\,\bibnamefont {Johnson}},
		\bibinfo {author} {\bibfnamefont {J.}\,\bibnamefont {Lewis}}, \bibinfo
		{author} {\bibfnamefont {M.}\,\bibnamefont {Raff}}, \bibinfo {author}
		{\bibfnamefont {K.}\,\bibnamefont {Roberts}},\ and\ \bibinfo {author}
		{\bibfnamefont {P.}\,\bibnamefont {Walter}},\ }\href@noop {} {\emph {\bibinfo
			{title} {Molecular Biology of the Cell}}},\ \bibinfo {edition} {6th}\ ed.\
	(\bibinfo  {publisher} {Garland Science},\ \bibinfo {address} {New York},\
	\bibinfo {year} {2014})\BibitemShut {NoStop}%
	\bibitem [{\citenamefont {Howard}(2001)}]{Howard2001}%
	\BibitemOpen
	\bibfield  {author} {\bibinfo {author} {\bibfnamefont {J.}\,\bibnamefont
			{Howard}},\ }\href@noop {} {\emph {\bibinfo {title} {Mechanics of Motor
				Proteins and the Cytoskeleton}}}\ (\bibinfo  {publisher} {Sinauer
		Associates},\ \bibinfo {address} {Sunderland, MA},\ \bibinfo {year}
	{2001})\BibitemShut {NoStop}%
	\bibitem [{\citenamefont {Reimann}(2002)}]{Reimann2002}%
	\BibitemOpen
	\bibfield  {author} {\bibinfo {author} {\bibfnamefont {P.}\,\bibnamefont
			{Reimann}},\ }\bibfield  {title} {\bibinfo {title} {Brownian motors: noisy
			transport far from equilibrium},\ }\href@noop {} {\bibfield  {journal}
		{\bibinfo  {journal} {Phys. Rep.}\ }\textbf {\bibinfo {volume}
			{\textbf{361}}},\ \bibinfo {pages} {57} (\bibinfo {year} {2002})}\BibitemShut
	{NoStop}%
	\bibitem [{\citenamefont {Astumian}(1997)}]{Astumian1997}%
	\BibitemOpen
	\bibfield  {author} {\bibinfo {author} {\bibfnamefont {R.\,D.}\ \bibnamefont
			{Astumian}},\ }\bibfield  {title} {\bibinfo {title} {Thermodynamics and
			kinetics of a brownian motor},\ }\href
	{https://doi.org/10.1126/science.276.5314.917} {\bibfield  {journal}
		{\bibinfo  {journal} {Science}\ }\textbf {\bibinfo {volume} {276}},\ \bibinfo
		{pages} {917} (\bibinfo {year} {1997})}\BibitemShut {NoStop}%
	\bibitem [{\citenamefont {Berg}\ and\ \citenamefont
		{Anderson}(1973)}]{BergAnderson1973}%
	\BibitemOpen
	\bibfield  {author} {\bibinfo {author} {\bibfnamefont {H.\,C.}\ \bibnamefont
			{Berg}}\ and\ \bibinfo {author} {\bibfnamefont {R.\,A.}\ \bibnamefont
			{Anderson}},\ }\bibfield  {title} {\bibinfo {title} {Bacteria swim by
			rotating their flagellar filaments},\ }\href
	{https://doi.org/10.1038/245380a0} {\bibfield  {journal} {\bibinfo  {journal}
			{Nature}\ }\textbf {\bibinfo {volume} {245}},\ \bibinfo {pages} {380}
		(\bibinfo {year} {1973})}\BibitemShut {NoStop}%
	\bibitem [{\citenamefont {Minamino}\ and\ \citenamefont
		{Imada}(2015)}]{MinaminoImada2015}%
	\BibitemOpen
	\bibfield  {author} {\bibinfo {author} {\bibfnamefont {T.}\,\bibnamefont
			{Minamino}}\ and\ \bibinfo {author} {\bibfnamefont {K.}\,\bibnamefont
			{Imada}},\ }\bibfield  {title} {\bibinfo {title} {The bacterial flagellar
			motor and its structural diversity},\ }\href
	{https://doi.org/10.1016/j.tim.2015.01.008} {\bibfield  {journal} {\bibinfo
			{journal} {Trends in Microbiology}\ }\textbf {\bibinfo {volume} {23}},\
		\bibinfo {pages} {267} (\bibinfo {year} {2015})}\BibitemShut {NoStop}%
	\bibitem [{\citenamefont {Johnson}\ \emph
		{et\,al.}(2021{\natexlab{a}})\citenamefont {Johnson}, \citenamefont {Furlong},
		\citenamefont {Deme}, \citenamefont {Nord}, \citenamefont {Caesar},
		\citenamefont {Chevance}, \citenamefont {Berry}, \citenamefont {Hughes},\
		and\ \citenamefont {Lea}}]{Johnson2021}%
	\BibitemOpen
	\bibfield  {author} {\bibinfo {author} {\bibfnamefont {S.}\,\bibnamefont
			{Johnson}}, \bibinfo {author} {\bibfnamefont {E.\,J.}\ \bibnamefont
			{Furlong}}, \bibinfo {author} {\bibfnamefont {J.\,C.}\ \bibnamefont {Deme}},
		\bibinfo {author} {\bibfnamefont {A.\,L.}\ \bibnamefont {Nord}}, \bibinfo
		{author} {\bibfnamefont {J.}\,\bibnamefont {Caesar}}, \bibinfo {author}
		{\bibfnamefont {F.\,F.\,V.}\ \bibnamefont {Chevance}}, \bibinfo {author}
		{\bibfnamefont {R.\,M.}\ \bibnamefont {Berry}}, \bibinfo {author}
		{\bibfnamefont {K.\,T.}\ \bibnamefont {Hughes}},\ and\ \bibinfo {author}
		{\bibfnamefont {S.\,M.}\ \bibnamefont {Lea}},\ }\bibfield  {title} {\bibinfo
		{title} {Structural basis of directional switching by the bacterial
			flagellum},\ }\href {https://doi.org/10.1038/s41586-020-03080-5} {\bibfield
		{journal} {\bibinfo  {journal} {Nature}\ }\textbf {\bibinfo {volume} {590}},\
		\bibinfo {pages} {233} (\bibinfo {year} {2021}{\natexlab{a}})}\BibitemShut
	{NoStop}%
	\bibitem [{\citenamefont {Yamaguchi}\ \emph {et\,al.}(2021)\citenamefont
		{Yamaguchi}, \citenamefont {Makino}, \citenamefont {Miyata}, \citenamefont
		{Minamino}, \citenamefont {Kato},\ and\ \citenamefont
		{Namba}}]{Yamaguchi2021}%
	\BibitemOpen
	\bibfield  {author} {\bibinfo {author} {\bibfnamefont {T.}\,\bibnamefont
			{Yamaguchi}}, \bibinfo {author} {\bibfnamefont {F.}\,\bibnamefont {Makino}},
		\bibinfo {author} {\bibfnamefont {T.}\,\bibnamefont {Miyata}}, \bibinfo
		{author} {\bibfnamefont {T.}\,\bibnamefont {Minamino}}, \bibinfo {author}
		{\bibfnamefont {T.}\,\bibnamefont {Kato}},\ and\ \bibinfo {author}
		{\bibfnamefont {K.}\,\bibnamefont {Namba}},\ }\bibfield  {title} {\bibinfo
		{title} {Structure of the molecular bushing of the bacterial flagellar
			motor},\ }\href {https://doi.org/10.1038/s41467-021-24715-3} {\bibfield
		{journal} {\bibinfo  {journal} {Nature Communications}\ }\textbf {\bibinfo
			{volume} {12}} (\bibinfo {year} {2021})}\BibitemShut {NoStop}%
	\bibitem [{\citenamefont {Kojima}\ and\ \citenamefont
		{Blair}(2004)}]{KojimaBlair2004}%
	\BibitemOpen
	\bibfield  {author} {\bibinfo {author} {\bibfnamefont {S.}\,\bibnamefont
			{Kojima}}\ and\ \bibinfo {author} {\bibfnamefont {D.\,F.}\ \bibnamefont
			{Blair}},\ }\bibfield  {title} {\bibinfo {title} {The bacterial flagellar
			motor: Structure and function of a complex molecular machine},\ }\href
	{https://doi.org/10.1016/S0074-7696(04)33003-1} {\bibfield  {journal}
		{\bibinfo  {journal} {International Review of Cytology}\ }\textbf {\bibinfo
			{volume} {233}},\ \bibinfo {pages} {93} (\bibinfo {year} {2004})}\BibitemShut
	{NoStop}%
	\bibitem [{\citenamefont {Sowa}\ \emph {et\,al.}(2005)\citenamefont {Sowa},
		\citenamefont {Rowe}, \citenamefont {Leake}, \citenamefont {Yakushi},
		\citenamefont {Homma}, \citenamefont {Ishijima},\ and\ \citenamefont
		{Berry}}]{SowaRoweLeakeYakushiHommaIshijimaBerry2005}%
	\BibitemOpen
	\bibfield  {author} {\bibinfo {author} {\bibfnamefont {Y.}\,\bibnamefont
			{Sowa}}, \bibinfo {author} {\bibfnamefont {A.\,D.}\ \bibnamefont {Rowe}},
		\bibinfo {author} {\bibfnamefont {M.\,C.}\ \bibnamefont {Leake}}, \bibinfo
		{author} {\bibfnamefont {T.}\,\bibnamefont {Yakushi}}, \bibinfo {author}
		{\bibfnamefont {M.}\,\bibnamefont {Homma}}, \bibinfo {author} {\bibfnamefont
			{A.}\,\bibnamefont {Ishijima}},\ and\ \bibinfo {author} {\bibfnamefont
			{R.\,M.}\ \bibnamefont {Berry}},\ }\bibfield  {title} {\bibinfo {title}
		{Direct observation of steps in rotation of the bacterial flagellar motor},\
	}\href {https://doi.org/10.1038/nature04003} {\bibfield  {journal} {\bibinfo
			{journal} {Nature}\ }\textbf {\bibinfo {volume} {437}},\ \bibinfo {pages}
		{916} (\bibinfo {year} {2005})}\BibitemShut {NoStop}%
	\bibitem [{\citenamefont {Nakamura}\ \emph {et\,al.}(2010)\citenamefont
		{Nakamura}, \citenamefont {Kami-ike}, \citenamefont {ichi P.\,Yokota},
		\citenamefont {Minamino},\ and\ \citenamefont {Namba}}]{Nakamura2010}%
	\BibitemOpen
	\bibfield  {author} {\bibinfo {author} {\bibfnamefont {S.}\,\bibnamefont
			{Nakamura}}, \bibinfo {author} {\bibfnamefont {N.}\,\bibnamefont {Kami-ike}},
		\bibinfo {author} {\bibfnamefont {J.}\,\bibnamefont {ichi P.\,Yokota}},
		\bibinfo {author} {\bibfnamefont {T.}\,\bibnamefont {Minamino}},\ and\
		\bibinfo {author} {\bibfnamefont {K.}\,\bibnamefont {Namba}},\ }\bibfield
	{title} {\bibinfo {title} {Evidence for symmetry in the elementary process of
			bidirectional torque generation by the bacterial flagellar motor},\
	}\href@noop {} {\bibfield  {journal} {\bibinfo  {journal} {Proceedings of the
				National Academy of Sciences}\ }\textbf {\bibinfo {volume} {107}},\ \bibinfo
		{pages} {17616} (\bibinfo {year} {2010})}\BibitemShut {NoStop}%
	\bibitem [{\citenamefont {Johnson}\ \emph
		{et\,al.}(2021{\natexlab{b}})\citenamefont {Johnson}, \citenamefont {Furlong},
		\citenamefont {Deme}, \citenamefont {Nord}, \citenamefont {Caesar},
		\citenamefont {Chevance}, \citenamefont {Berry}, \citenamefont {Hughes},\
		and\ \citenamefont
		{Lea}}]{JohnsonFurlongDemeNordCaesarChevanceBerryHughesLea2021}%
	\BibitemOpen
	\bibfield  {author} {\bibinfo {author} {\bibfnamefont {S.}\,\bibnamefont
			{Johnson}}, \bibinfo {author} {\bibfnamefont {E.\,J.}\ \bibnamefont
			{Furlong}}, \bibinfo {author} {\bibfnamefont {J.\,C.}\ \bibnamefont {Deme}},
		\bibinfo {author} {\bibfnamefont {A.\,L.}\ \bibnamefont {Nord}}, \bibinfo
		{author} {\bibfnamefont {J.\,J.\,E.}\ \bibnamefont {Caesar}}, \bibinfo {author}
		{\bibfnamefont {F.\,F.\,V.}\ \bibnamefont {Chevance}}, \bibinfo {author}
		{\bibfnamefont {R.\,M.}\ \bibnamefont {Berry}}, \bibinfo {author}
		{\bibfnamefont {K.\,T.}\ \bibnamefont {Hughes}},\ and\ \bibinfo {author}
		{\bibfnamefont {S.\,M.}\ \bibnamefont {Lea}},\ }\bibfield  {title} {\bibinfo
		{title} {Molecular structure of the intact bacterial flagellar basal body},\
	}\href {https://doi.org/10.1038/s41564-021-00895-y} {\bibfield  {journal}
		{\bibinfo  {journal} {Nature Microbiology}\ }\textbf {\bibinfo {volume}
			{6}},\ \bibinfo {pages} {712} (\bibinfo {year}
		{2021}{\natexlab{b}})}\BibitemShut {NoStop}%
	\bibitem [{\citenamefont {Drobnič}\ \emph {et\,al.}(2025)\citenamefont
		{Drobnič}, \citenamefont {Cohen}, \citenamefont {Calcraft}, \citenamefont
		{Alzheimer}, \citenamefont {Froschauer}, \citenamefont {Svensson},
		\citenamefont {Hoffmann}, \citenamefont {Singh}, \citenamefont {Garg},
		\citenamefont {Henderson}, \citenamefont {Umrekar}, \citenamefont {Nans},
		\citenamefont {Ribardo}, \citenamefont {Pedaci}, \citenamefont {Nord},
		\citenamefont {Hochberg}, \citenamefont {Hendrixson}, \citenamefont {Sharma},
		\citenamefont {Rosenthal},\ and\ \citenamefont {Beeby}}]{Drobnic2025}%
	\BibitemOpen
	\bibfield  {author} {\bibinfo {author} {\bibfnamefont {T.}\,\bibnamefont
			{Drobnič}}, \bibinfo {author} {\bibfnamefont {E.}\,\bibnamefont {Cohen}},
		\bibinfo {author} {\bibfnamefont {T.}\,\bibnamefont {Calcraft}}, \bibinfo
		{author} {\bibfnamefont {M.}\,\bibnamefont {Alzheimer}}, \bibinfo {author}
		{\bibfnamefont {K.}\,\bibnamefont {Froschauer}}, \bibinfo {author}
		{\bibfnamefont {S.}\,\bibnamefont {Svensson}}, \bibinfo {author}
		{\bibfnamefont {W.}\,\bibnamefont {Hoffmann}}, \bibinfo {author}
		{\bibfnamefont {N.}\,\bibnamefont {Singh}}, \bibinfo {author} {\bibfnamefont
			{S.}\,\bibnamefont {Garg}}, \bibinfo {author} {\bibfnamefont {L.}\,\bibnamefont
			{Henderson}}, \bibinfo {author} {\bibfnamefont {T.}\,\bibnamefont {Umrekar}},
		\bibinfo {author} {\bibfnamefont {A.}\,\bibnamefont {Nans}}, \bibinfo {author}
		{\bibfnamefont {D.}\,\bibnamefont {Ribardo}}, \bibinfo {author} {\bibfnamefont
			{F.}\,\bibnamefont {Pedaci}}, \bibinfo {author} {\bibfnamefont
			{A.}\,\bibnamefont {Nord}}, \bibinfo {author} {\bibfnamefont {G.}\,\bibnamefont
			{Hochberg}}, \bibinfo {author} {\bibfnamefont {D.}\,\bibnamefont
			{Hendrixson}}, \bibinfo {author} {\bibfnamefont {C.}\,\bibnamefont {Sharma}},
		\bibinfo {author} {\bibfnamefont {P.}\,\bibnamefont {Rosenthal}},\ and\
		\bibinfo {author} {\bibfnamefont {M.}\,\bibnamefont {Beeby}},\ }\bibfield
	{title} {\bibinfo {title} {In situ structure of a bacterial flagellar motor
			at subnanometre resolution reveals adaptations for increased torque},\ }\href
	{https://doi.org/10.1038/s41564-025-02012-9} {\bibfield  {journal} {\bibinfo
			{journal} {Nature Microbiology}\ }\textbf {\bibinfo {volume} {10}},\ \bibinfo
		{pages} {1723} (\bibinfo {year} {2025})}\BibitemShut {NoStop}%
	\bibitem [{\citenamefont {Tan}\ \emph {et\,al.}(2024)\citenamefont {Tan},
		\citenamefont {Zhang}, \citenamefont {Zhou}, \citenamefont {Han},
		\citenamefont {Zhou}, \citenamefont {Zhu} \emph {et\,al.}}]{Tan2024}%
	\BibitemOpen
	\bibfield  {author} {\bibinfo {author} {\bibfnamefont {J.}\,\bibnamefont
			{Tan}}, \bibinfo {author} {\bibfnamefont {L.}\,\bibnamefont {Zhang}}, \bibinfo
		{author} {\bibfnamefont {X.}\,\bibnamefont {Zhou}}, \bibinfo {author}
		{\bibfnamefont {S.}\,\bibnamefont {Han}}, \bibinfo {author} {\bibfnamefont
			{Y.}\,\bibnamefont {Zhou}}, \bibinfo {author} {\bibfnamefont {Y.}\,\bibnamefont
			{Zhu}}, \emph {et\,al.},\ }\bibfield  {title} {\bibinfo {title} {Structural
			basis of the bacterial flagellar motor rotational switching},\ }\href
	{https://doi.org/10.1038/s41422-024-01017-z} {\bibfield  {journal} {\bibinfo
			{journal} {Cell Research}\ }\textbf {\bibinfo {volume} {34}},\ \bibinfo
		{pages} {788} (\bibinfo {year} {2024})}\BibitemShut {NoStop}%
	\bibitem [{\citenamefont {Nakamura}\ and\ \citenamefont
		{Minamino}(2024)}]{Nakamura2023}%
	\BibitemOpen
	\bibfield  {author} {\bibinfo {author} {\bibfnamefont {S.}\,\bibnamefont
			{Nakamura}}\ and\ \bibinfo {author} {\bibfnamefont {T.}\,\bibnamefont
			{Minamino}},\ }\bibfield  {title} {\bibinfo {title} {Structure and dynamics
			of the bacterial flagellar motor complex},\ }\href
	{https://doi.org/10.3390/biom14121488} {\bibfield  {journal} {\bibinfo
			{journal} {Biomolecules}\ }\textbf {\bibinfo {volume} {14}},\ \bibinfo
		{pages} {1488} (\bibinfo {year} {2024})}\BibitemShut {NoStop}%
	\bibitem [{\citenamefont {Xing}\ \emph {et\,al.}(2006)\citenamefont {Xing},
		\citenamefont {Bai}, \citenamefont {Berry},\ and\ \citenamefont
		{Oster}}]{Xing2006}%
	\BibitemOpen
	\bibfield  {author} {\bibinfo {author} {\bibfnamefont {J.}\,\bibnamefont
			{Xing}}, \bibinfo {author} {\bibfnamefont {F.}\,\bibnamefont {Bai}}, \bibinfo
		{author} {\bibfnamefont {R.}\,\bibnamefont {Berry}},\ and\ \bibinfo {author}
		{\bibfnamefont {G.}\,\bibnamefont {Oster}},\ }\bibfield  {title} {\bibinfo
		{title} {Torque-speed relationship of the bacterial flagellar motor},\
	}\href {https://doi.org/110.1073/pnas.0507959103} {\bibfield  {journal}
		{\bibinfo  {journal} {Proceedings of the National Academy of Sciences of the
				United States of America}\ }\textbf {\bibinfo {volume} {103}},\ \bibinfo
		{pages} {1260-1265} (\bibinfo {year} {2006})}\BibitemShut {NoStop}%
	\bibitem [{\citenamefont {Mora}\ \emph {et\,al.}(2009)\citenamefont {Mora},
		\citenamefont {Yu}, \citenamefont {Sowa},\ and\ \citenamefont
		{Wingreen}}]{MoraYuSowaWingreen2009}%
	\BibitemOpen
	\bibfield  {author} {\bibinfo {author} {\bibfnamefont {T.}\,\bibnamefont
			{Mora}}, \bibinfo {author} {\bibfnamefont {H.}\,\bibnamefont {Yu}}, \bibinfo
		{author} {\bibfnamefont {Y.}\,\bibnamefont {Sowa}},\ and\ \bibinfo {author}
		{\bibfnamefont {N.\,S.}\ \bibnamefont {Wingreen}},\ }\bibfield  {title}
	{\bibinfo {title} {Steps in the bacterial flagellar motor},\ }\href
	{https://journals.plos.org/ploscompbiol/article?id=10.1371/journal.pcbi.1000540}
	{\bibfield  {journal} {\bibinfo  {journal} {PLoS computational biology}\
		}\textbf {\bibinfo {volume} {5}},\ \bibinfo {pages} {e1000540} (\bibinfo
		{year} {2009})}\BibitemShut {NoStop}%
	\bibitem [{\citenamefont {Rieu}\ \emph {et\,al.}(2025)\citenamefont {Rieu},
		\citenamefont {Nord}, \citenamefont {Courbet}, \citenamefont {El\,Sayyed},\
		and\ \citenamefont {Berry}}]{RieuNordCourbetElSayyedEtAl2025}%
	\BibitemOpen
	\bibfield  {author} {\bibinfo {author} {\bibfnamefont {M.}\,\bibnamefont
			{Rieu}}, \bibinfo {author} {\bibfnamefont {A.\,L.}\ \bibnamefont {Nord}},
		\bibinfo {author} {\bibfnamefont {A.}\,\bibnamefont {Courbet}}, \bibinfo
		{author} {\bibfnamefont {H.}\,\bibnamefont {El\,Sayyed}},\ and\ \bibinfo
		{author} {\bibfnamefont {R.\,M.}\ \bibnamefont {Berry}},\ }\bibfield  {title}
	{\bibinfo {title} {Single-molecule observation of multi-scale dynamic
			heterogeneity in the molecular bearing of the bacterial flagellum},\
	}\bibfield  {journal} {\bibinfo  {journal} {bioRxiv}\ }\href
	{https://doi.org/10.1101/2025.02.20.639300} {10.1101/2025.02.20.639300}
	(\bibinfo {year} {2025})\BibitemShut {NoStop}%
	\bibitem [{\citenamefont {Hayashi}\ \emph {et\,al.}(2015)\citenamefont
		{Hayashi}, \citenamefont {Sasaki}, \citenamefont {Nakamura}, \citenamefont
		{Kudo}, \citenamefont {Inoue}, \citenamefont {Noji},\ and\ \citenamefont
		{Hayashi}}]{HayashiSasakiNakamuraKudoInoueNoji2015}%
	\BibitemOpen
	\bibfield  {author} {\bibinfo {author} {\bibfnamefont {R.}\,\bibnamefont
			{Hayashi}}, \bibinfo {author} {\bibfnamefont {K.}\,\bibnamefont {Sasaki}},
		\bibinfo {author} {\bibfnamefont {S.}\,\bibnamefont {Nakamura}}, \bibinfo
		{author} {\bibfnamefont {S.}\,\bibnamefont {Kudo}}, \bibinfo {author}
		{\bibfnamefont {Y.}\,\bibnamefont {Inoue}}, \bibinfo {author} {\bibfnamefont
			{H.}\,\bibnamefont {Noji}},\ and\ \bibinfo {author} {\bibfnamefont
			{K.}\,\bibnamefont {Hayashi}},\ }\bibfield  {title} {\bibinfo {title} {Giant
			acceleration of diffusion observed in a single-molecule experiment on
			{F}$_1$-{A}{T}{P}ase},\ }\href
	{https://journals.aps.org/prl/abstract/10.1103/PhysRevLett.114.248101}
	{\bibfield  {journal} {\bibinfo  {journal} {Phys. Rev. Lett.}\ }\textbf
		{\bibinfo {volume} {\textbf{114}}},\ \bibinfo {pages} {248101} (\bibinfo
		{year} {2015})}\BibitemShut {NoStop}%
	\bibitem [{\citenamefont {Berry}\ and\ \citenamefont {Berg}(1997)}]{Berry1997}%
	\BibitemOpen
	\bibfield  {author} {\bibinfo {author} {\bibfnamefont {R.\,M.}\ \bibnamefont
			{Berry}}\ and\ \bibinfo {author} {\bibfnamefont {H.\,C.}\ \bibnamefont
			{Berg}},\ }\bibfield  {title} {\bibinfo {title} {Absence of a barrier to
			backwards rotation of the bacterial flagellar motor demonstrated with optical
			tweezers},\ }\href {https://doi.org/10.1073/pnas.94.26.14433} {\bibfield
		{journal} {\bibinfo  {journal} {Proceedings of the National Academy of
				Sciences of the United States of America}\ }\textbf {\bibinfo {volume}
			{94}},\ \bibinfo {pages} {14433} (\bibinfo {year} {1997})}\BibitemShut
	{NoStop}%
	\bibitem [{\citenamefont {Wang}\ \emph {et\,al.}(2022)\citenamefont {Wang},
		\citenamefont {Yue}, \citenamefont {Zhang},\ and\ \citenamefont
		{Yuan}}]{Wang2022}%
	\BibitemOpen
	\bibfield  {author} {\bibinfo {author} {\bibfnamefont {B.}\,\bibnamefont
			{Wang}}, \bibinfo {author} {\bibfnamefont {G.}\,\bibnamefont {Yue}}, \bibinfo
		{author} {\bibfnamefont {R.}\,\bibnamefont {Zhang}},\ and\ \bibinfo {author}
		{\bibfnamefont {J.}\,\bibnamefont {Yuan}},\ }\bibfield  {title} {\bibinfo
		{title} {Direct measurement of the stall torque of the flagellar motor in
			{E}scherichia coli with magnetic tweezers},\ }\href
	{https://doi.org/10.1128/mbio.00782-22} {\bibfield  {journal} {\bibinfo
			{journal} {mBio}\ }\textbf {\bibinfo {volume} {13}},\ \bibinfo {pages}
		{e00782} (\bibinfo {year} {2022})}\BibitemShut {NoStop}%
	\bibitem [{\citenamefont {Berry}\ \emph {et\,al.}(1995)\citenamefont {Berry},
		\citenamefont {Turner},\ and\ \citenamefont {Berg}}]{Berry1995}%
	\BibitemOpen
	\bibfield  {author} {\bibinfo {author} {\bibfnamefont {R.\,M.}\ \bibnamefont
			{Berry}}, \bibinfo {author} {\bibfnamefont {L.}\,\bibnamefont {Turner}},\ and\
		\bibinfo {author} {\bibfnamefont {H.\,C.}\ \bibnamefont {Berg}},\ }\bibfield
	{title} {\bibinfo {title} {Mechanical limits of bacterial flagellar motors
			probed by electrorotation},\ }\href
	{https://doi.org/10.1016/S0006-3495(95)79900-3} {\bibfield  {journal}
		{\bibinfo  {journal} {Biophysical Journal}\ }\textbf {\bibinfo {volume}
			{69}},\ \bibinfo {pages} {280} (\bibinfo {year} {1995})}\BibitemShut
	{NoStop}%
	\bibitem [{\citenamefont {Hoffmann}\ \emph {et\,al.}(2025)\citenamefont
		{Hoffmann}, \citenamefont {Biquet-Bisquert}, \citenamefont {Pedaci},\ and\
		\citenamefont {Nord}}]{Hoffmann2025}%
	\BibitemOpen
	\bibfield  {author} {\bibinfo {author} {\bibfnamefont {W.\,H.}\ \bibnamefont
			{Hoffmann}}, \bibinfo {author} {\bibfnamefont {A.}\,\bibnamefont
			{Biquet-Bisquert}}, \bibinfo {author} {\bibfnamefont {F.}\,\bibnamefont
			{Pedaci}},\ and\ \bibinfo {author} {\bibfnamefont {A.\,L.}\ \bibnamefont
			{Nord}},\ }\bibfield  {title} {\bibinfo {title} {Measuring bacterial
			flagellar motor dynamics via a bead assay},\ }in\ \href
	{https://doi.org/10.1007/978-1-0716-4280-1_2} {\emph {\bibinfo {booktitle}
			{Molecular Motors: Methods and Protocols}}},\ Vol.\ \bibinfo {volume}
	{2881},\ \bibinfo {editor} {edited by\ \bibinfo {editor} {\bibfnamefont
			{C.}\,\bibnamefont {Lavelle}}\ and\ \bibinfo {editor} {\bibfnamefont
			{A.}\,\bibnamefont {Le\,Gall}}}\ (\bibinfo  {publisher} {Humana Press},\
	\bibinfo {year} {2025})\ pp.\ \bibinfo {pages} {43--64}\BibitemShut {NoStop}%
	\bibitem [{\citenamefont {Risken}(1996)}]{Risken1996}%
	\BibitemOpen
	\bibfield  {author} {\bibinfo {author} {\bibfnamefont {H.}\,\bibnamefont
			{Risken}},\ }\href {https://link.springer.com/book/10.1007/978-3-642-61544-3}
	{\emph {\bibinfo {title} {Fokker-{P}lanck equation}}}\ (\bibinfo  {publisher}
	{Springer},\ \bibinfo {year} {1996})\BibitemShut {NoStop}%
	\bibitem [{\citenamefont {Gardiner}(2009)}]{Gardiner2009}%
	\BibitemOpen
	\bibfield  {author} {\bibinfo {author} {\bibfnamefont {C.\,W.}\ \bibnamefont
			{Gardiner}},\ }\href {https://link.springer.com/book/9783540707127} {\emph
		{\bibinfo {title} {Handbook of Stochastic Methods for Physics, Chemistry and
				the Natural Sciences}}},\ \bibinfo {edition} {4th}\ ed.\ (\bibinfo
	{publisher} {Springer},\ \bibinfo {year} {2009})\BibitemShut {NoStop}%
	\bibitem [{\citenamefont {Reimann}\ \emph {et\,al.}(2001)\citenamefont
		{Reimann}, \citenamefont {{Van den Broeck}}, \citenamefont {Linke},
		\citenamefont {H\"anggi}, \citenamefont {Rubi},\ and\ \citenamefont
		{P\'erez-Madrid}}]{ReimannVandenBroeckLinkeEtAl2001}%
	\BibitemOpen
	\bibfield  {author} {\bibinfo {author} {\bibfnamefont {P.}\,\bibnamefont
			{Reimann}}, \bibinfo {author} {\bibfnamefont {C.}\,\bibnamefont {{Van den
					Broeck}}}, \bibinfo {author} {\bibfnamefont {H.}\,\bibnamefont {Linke}},
		\bibinfo {author} {\bibfnamefont {P.}\,\bibnamefont {H\"anggi}}, \bibinfo
		{author} {\bibfnamefont {J.\,M.}\ \bibnamefont {Rubi}},\ and\ \bibinfo
		{author} {\bibfnamefont {A.}\,\bibnamefont {P\'erez-Madrid}},\ }\bibfield
	{title} {\bibinfo {title} {Giant acceleration of free diffusion by use of
			tilted periodic potentials},\ }\href
	{https://journals.aps.org/prl/abstract/10.1103/PhysRevLett.87.010602}
	{\bibfield  {journal} {\bibinfo  {journal} {Phys. Rev. Lett.}\ }\textbf
		{\bibinfo {volume} {\textbf{87}}},\ \bibinfo {pages} {010602} (\bibinfo
		{year} {2001})}\BibitemShut {NoStop}%
	\bibitem [{\citenamefont {Barato}\ and\ \citenamefont
		{Seifert}(2015)}]{BaratoSeifert2015}%
	\BibitemOpen
	\bibfield  {author} {\bibinfo {author} {\bibfnamefont {A.\,C.}\ \bibnamefont
			{Barato}}\ and\ \bibinfo {author} {\bibfnamefont {U.}\,\bibnamefont
			{Seifert}},\ }\bibfield  {title} {\bibinfo {title} {Thermodynamic uncertainty
			relation for biomolecular processes},\ }\href
	{https://doi.org/10.1103/PhysRevLett.114.158101} {\bibfield  {journal}
		{\bibinfo  {journal} {Phys. Rev. Lett.}\ }\textbf {\bibinfo {volume} {114}},\
		\bibinfo {pages} {158101} (\bibinfo {year} {2015})}\BibitemShut {NoStop}%
	\bibitem [{\citenamefont {Nord}\ and\ \citenamefont
		{Pedaci}(2020)}]{NordPedaci2020}%
	\BibitemOpen
	\bibfield  {author} {\bibinfo {author} {\bibfnamefont {A.\,L.}\ \bibnamefont
			{Nord}}\ and\ \bibinfo {author} {\bibfnamefont {F.}\,\bibnamefont {Pedaci}},\
	}\bibinfo {title} {Mechanisms and dynamics of the bacterial flagellar
		motor},\ in\ \href {https://doi.org/10.1007/978-3-030-46886-6_5} {\emph
		{\bibinfo {booktitle} {Physical Microbiology}}},\ \bibinfo {editor} {edited
		by\ \bibinfo {editor} {\bibfnamefont {G.}\,\bibnamefont {Dum{\'e}nil}}\ and\
		\bibinfo {editor} {\bibfnamefont {S.}\,\bibnamefont {van Teeffelen}}}\
	(\bibinfo  {publisher} {Springer International Publishing},\ \bibinfo
	{address} {Cham},\ \bibinfo {year} {2020})\ pp.\ \bibinfo {pages}
	{81--100}\BibitemShut {NoStop}%
	\bibitem [{\citenamefont {Wang}\ \emph {et\,al.}(2020)\citenamefont {Wang},
		\citenamefont {Chen}, \citenamefont {Zhang},\ and\ \citenamefont
		{Yuan}}]{WangChenZhangEtAl2020}%
	\BibitemOpen
	\bibfield  {author} {\bibinfo {author} {\bibfnamefont {R.}\,\bibnamefont
			{Wang}}, \bibinfo {author} {\bibfnamefont {Q.}\,\bibnamefont {Chen}}, \bibinfo
		{author} {\bibfnamefont {R.}\,\bibnamefont {Zhang}},\ and\ \bibinfo {author}
		{\bibfnamefont {J.}\,\bibnamefont {Yuan}},\ }\bibfield  {title} {\bibinfo
		{title} {Measurement of the internal frictional drag of the bacterial
			flagellar motor by fluctuation analysis},\ }\href
	{https://doi.org/https://doi.org/10.1016/j.bpj.2020.04.020} {\bibfield
		{journal} {\bibinfo  {journal} {Biophysical Journal}\ }\textbf {\bibinfo
			{volume} {118}},\ \bibinfo {pages} {2718} (\bibinfo {year}
		{2020})}\BibitemShut {NoStop}%
	\bibitem [{\citenamefont {Challis}\ and\ \citenamefont
		{Jack}(2013)}]{ChallisJack2013}%
	\BibitemOpen
	\bibfield  {author} {\bibinfo {author} {\bibfnamefont {K.\,J.}\ \bibnamefont
			{Challis}}\ and\ \bibinfo {author} {\bibfnamefont {M.\,W.}\ \bibnamefont
			{Jack}},\ }\bibfield  {title} {\bibinfo {title} {Energy transfer in a
			molecular motor in the kramers regime},\ }\href
	{https://doi.org/10.1103/PhysRevE.88.042114} {\bibfield  {journal} {\bibinfo
			{journal} {Phys. Rev. E}\ }\textbf {\bibinfo {volume} {88}},\ \bibinfo
		{pages} {042114} (\bibinfo {year} {2013})}\BibitemShut {NoStop}%
	\bibitem [{\citenamefont {Challis}(2016)}]{Challis2016}%
	\BibitemOpen
	\bibfield  {author} {\bibinfo {author} {\bibfnamefont {K.\,J.}\ \bibnamefont
			{Challis}},\ }\bibfield  {title} {\bibinfo {title} {Numerical study of the
			tight-binding approach to overdamped brownian motion on a tilted periodic
			potential},\ }\href {https://doi.org/10.1103/PhysRevE.94.062123} {\bibfield
		{journal} {\bibinfo  {journal} {Phys. Rev. E}\ }\textbf {\bibinfo {volume}
			{94}},\ \bibinfo {pages} {062123} (\bibinfo {year} {2016})}\BibitemShut
	{NoStop}%
	\bibitem [{\citenamefont {L\'opez-Alamilla}\ \emph {et\,al.}(2018)\citenamefont
		{L\'opez-Alamilla}, \citenamefont {Jack},\ and\ \citenamefont
		{Challis}}]{Lopez-AlamillaJackChallis2018}%
	\BibitemOpen
	\bibfield  {author} {\bibinfo {author} {\bibfnamefont {N.\,J.}\ \bibnamefont
			{L\'opez-Alamilla}}, \bibinfo {author} {\bibfnamefont {M.\,W.}\ \bibnamefont
			{Jack}},\ and\ \bibinfo {author} {\bibfnamefont {K.\,J.}\ \bibnamefont
			{Challis}},\ }\bibfield  {title} {\bibinfo {title} {Reconstructing
			free-energy landscapes for nonequilibrium periodic potentials},\ }\href
	{https://doi.org/doi.org/10.1103/PhysRevE.97.032419} {\bibfield  {journal}
		{\bibinfo  {journal} {Phys. Rev. E}\ }\textbf {\bibinfo {volume}
			{\textbf{97}}},\ \bibinfo {pages} {032419} (\bibinfo {year}
		{2018})}\BibitemShut {NoStop}%
	\bibitem [{\citenamefont {L\'{o}pez-Alamilla}\ \emph
		{et\,al.}(2020)\citenamefont {L\'{o}pez-Alamilla}, \citenamefont {Jack},\ and\
		\citenamefont {Challis}}]{Lopez-AlamillaJackChallis2020}%
	\BibitemOpen
	\bibfield  {author} {\bibinfo {author} {\bibfnamefont {N.\,J.}\ \bibnamefont
			{L\'{o}pez-Alamilla}}, \bibinfo {author} {\bibfnamefont {M.\,W.}\ \bibnamefont
			{Jack}},\ and\ \bibinfo {author} {\bibfnamefont {K.\,J.}\ \bibnamefont
			{Challis}},\ }\bibfield  {title} {\bibinfo {title} {Enhanced diffusion and
			the eigenvalue band structure of brownian motion in tilted periodic
			potentials},\ }\href {https://doi.org/10.1103/PhysRevE.102.042405} {\bibfield
		{journal} {\bibinfo  {journal} {Phys. Rev. E}\ }\textbf {\bibinfo {volume}
			{\textbf{102}}},\ \bibinfo {pages} {042405} (\bibinfo {year}
		{2020})}\BibitemShut {NoStop}%
	\bibitem [{\citenamefont {Jack}\ \emph {et\,al.}(2020)\citenamefont {Jack},
		\citenamefont {L\'opez-Alamilla},\ and\ \citenamefont
		{Challis}}]{JackLopez-AlamillaChallis2020}%
	\BibitemOpen
	\bibfield  {author} {\bibinfo {author} {\bibfnamefont {M.\,W.}\ \bibnamefont
			{Jack}}, \bibinfo {author} {\bibfnamefont {N.\,J.}\ \bibnamefont
			{L\'opez-Alamilla}},\ and\ \bibinfo {author} {\bibfnamefont {K.\,J.}\
			\bibnamefont {Challis}},\ }\bibfield  {title} {\bibinfo {title}
		{Thermodynamic uncertainty relations and molecular-scale energy conversion},\
	}\href {https://doi.org/doi.org/10.1103/PhysRevE.101.062123} {\bibfield
		{journal} {\bibinfo  {journal} {Phys. Rev. E}\ }\textbf {\bibinfo {volume}
			{\textbf{101}}},\ \bibinfo {pages} {062123} (\bibinfo {year}
		{2020})}\BibitemShut {NoStop}%
	\bibitem [{\citenamefont {Kittel}(2004)}]{Kittel2004}%
	\BibitemOpen
	\bibfield  {author} {\bibinfo {author} {\bibfnamefont {C.}\,\bibnamefont
			{Kittel}},\ }\href
	{https://www.wiley.com/en-us/Introduction+to+Solid+State+Physics%2C+8th+Edition-p-9780471415268}
	{\emph {\bibinfo {title} {Introduction to Solid State Physics}}}\ (\bibinfo
	{publisher} {Wiley},\ \bibinfo {year} {2004})\BibitemShut {NoStop}%
	\bibitem [{\citenamefont {Festa}\ and\ \citenamefont
		{d'Agliano}(1978)}]{Festa1978}%
	\BibitemOpen
	\bibfield  {author} {\bibinfo {author} {\bibfnamefont {R.}\,\bibnamefont
			{Festa}}\ and\ \bibinfo {author} {\bibfnamefont {E.}\,\bibnamefont
			{d'Agliano}},\ }\bibfield  {title} {\bibinfo {title} {Diffusion coefficient
			for a brownian particle in a periodic field of force: I. large friction
			limit},\ }\href
	{https://doi.org/https://doi.org/10.1016/0378-4371(78)90111-5} {\bibfield
		{journal} {\bibinfo  {journal} {Physica A}\ }\textbf {\bibinfo {volume}
			{90}},\ \bibinfo {pages} {229 } (\bibinfo {year} {1978})}\BibitemShut
	{NoStop}%
	\bibitem [{\citenamefont {L\'opez-Alamilla}\ and\ \citenamefont
		{Cachi}(2022)}]{Lopez-AlamillaCachi2022a}%
	\BibitemOpen
	\bibfield  {author} {\bibinfo {author} {\bibfnamefont {N.\,J.}\ \bibnamefont
			{L\'opez-Alamilla}}\ and\ \bibinfo {author} {\bibfnamefont {R.}\,\bibnamefont
			{Cachi}},\ }\bibfield  {title} {\bibinfo {title} {Virial-like thermodynamic
			uncertainty relation in the tight-binding regime},\ }\href
	{https://doi.org/doi.org/10.1063/5.0107554} {\bibfield  {journal} {\bibinfo
			{journal} {AIP Chaos}\ }\textbf {\bibinfo {volume} {\textbf{32}}},\ \bibinfo
		{pages} {103109} (\bibinfo {year} {2022})}\BibitemShut {NoStop}%
	\bibitem [{\citenamefont {Nord}\ \emph {et\,al.}(2022)\citenamefont {Nord},
		\citenamefont {Biquet-Bisquert}, \citenamefont {Abkarian}, \citenamefont
		{Pigaglio}, \citenamefont {Seduk}, \citenamefont {Magalon},\ and\
		\citenamefont {Pedaci}}]{NordBiquet-BisquertAbkarianPigaglioEtAl2022}%
	\BibitemOpen
	\bibfield  {author} {\bibinfo {author} {\bibfnamefont {A.\,L.}\ \bibnamefont
			{Nord}}, \bibinfo {author} {\bibfnamefont {A.}\,\bibnamefont
			{Biquet-Bisquert}}, \bibinfo {author} {\bibfnamefont {M.}\,\bibnamefont
			{Abkarian}}, \bibinfo {author} {\bibfnamefont {T.}\,\bibnamefont {Pigaglio}},
		\bibinfo {author} {\bibfnamefont {F.}\,\bibnamefont {Seduk}}, \bibinfo
		{author} {\bibfnamefont {A.}\,\bibnamefont {Magalon}},\ and\ \bibinfo {author}
		{\bibfnamefont {F.}\,\bibnamefont {Pedaci}},\ }\bibfield  {title} {\bibinfo
		{title} {Dynamic stiffening of the flagellar hook},\ }\href
	{https://doi.org/10.1038/s41467-022-30295-7} {\bibfield  {journal} {\bibinfo
			{journal} {Nature Communications}\ }\textbf {\bibinfo {volume} {13}},\
		\bibinfo {pages} {2925} (\bibinfo {year} {2022})}\BibitemShut {NoStop}%
	\bibitem [{\citenamefont {L\'opez-Alamilla}\ \emph
		{et\,al.}(2019{\natexlab{a}})\citenamefont {L\'opez-Alamilla}, \citenamefont
		{Jack},\ and\ \citenamefont {Challis}}]{Lopez-AlamillaJackChallis2019a}%
	\BibitemOpen
	\bibfield  {author} {\bibinfo {author} {\bibfnamefont {N.\,J.}\ \bibnamefont
			{L\'opez-Alamilla}}, \bibinfo {author} {\bibfnamefont {M.\,W.}\ \bibnamefont
			{Jack}},\ and\ \bibinfo {author} {\bibfnamefont {K.\,J.}\ \bibnamefont
			{Challis}},\ }\bibfield  {title} {\bibinfo {title} {Analysing single-molecule
			trajectories to reconstruct free-energy landscapes of cyclic motor
			proteins},\ }\href {https://doi.org/doi.org/10.1016/j.jtbi.2018.11.015}
	{\bibfield  {journal} {\bibinfo  {journal} {J. Theor. Biol.}\ }\textbf
		{\bibinfo {volume} {\textbf{462}}},\ \bibinfo {pages} {321} (\bibinfo {year}
		{2019}{\natexlab{a}})}\BibitemShut {NoStop}%
	\bibitem [{\citenamefont {L\'opez-Alamilla}\ \emph
		{et\,al.}(2019{\natexlab{b}})\citenamefont {L\'opez-Alamilla}, \citenamefont
		{Jack},\ and\ \citenamefont {Challis}}]{Lopez-AlamillaJackChallis2019}%
	\BibitemOpen
	\bibfield  {author} {\bibinfo {author} {\bibfnamefont {N.\,J.}\ \bibnamefont
			{L\'opez-Alamilla}}, \bibinfo {author} {\bibfnamefont {M.\,W.}\ \bibnamefont
			{Jack}},\ and\ \bibinfo {author} {\bibfnamefont {K.\,J.}\ \bibnamefont
			{Challis}},\ }\bibfield  {title} {\bibinfo {title} {Reconstructing
			free-energy landscapes for cyclic molecular motors using full
			multidimensional or partial one dimensional dynamic information},\ }\href
	{https://doi.org/doi.org/10.1103/PhysRevE.100.012404} {\bibfield  {journal}
		{\bibinfo  {journal} {Phys. Rev. E}\ }\textbf {\bibinfo {volume}
			{\textbf{100}}},\ \bibinfo {pages} {012404} (\bibinfo {year}
		{2019}{\natexlab{b}})}\BibitemShut {NoStop}%
	\bibitem [{\citenamefont {Nord}\ \emph {et\,al.}(2017)\citenamefont {Nord},
		\citenamefont {Gachon}, \citenamefont {Perez-Carrasco}, \citenamefont
		{Nirody}, \citenamefont {Barducci}, \citenamefont {Berry},\ and\
		\citenamefont {Pedaci}}]{Nord2017}%
	\BibitemOpen
	\bibfield  {author} {\bibinfo {author} {\bibfnamefont {A.\,L.}\ \bibnamefont
			{Nord}}, \bibinfo {author} {\bibfnamefont {E.}\,\bibnamefont {Gachon}},
		\bibinfo {author} {\bibfnamefont {R.}\,\bibnamefont {Perez-Carrasco}},
		\bibinfo {author} {\bibfnamefont {J.\,A.}\ \bibnamefont {Nirody}}, \bibinfo
		{author} {\bibfnamefont {A.}\,\bibnamefont {Barducci}}, \bibinfo {author}
		{\bibfnamefont {R.\,M.}\ \bibnamefont {Berry}},\ and\ \bibinfo {author}
		{\bibfnamefont {F.}\,\bibnamefont {Pedaci}},\ }\bibfield  {title} {\bibinfo
		{title} {{Catch bond drives stator mechanosensitivity in the bacterial
				flagellar motor}},\ }\href {https://doi.org/10.1073/pnas.1716002114}
	{\bibfield  {journal} {\bibinfo  {journal} {Proceedings of the National
				Academy of Sciences}\ }\textbf {\bibinfo {volume} {114}},\ \bibinfo {pages}
		{12952} (\bibinfo {year} {2017})}\BibitemShut {NoStop}%
	\bibitem [{\citenamefont {Leake}\ \emph {et\,al.}(2006)\citenamefont {Leake},
		\citenamefont {Chandler}, \citenamefont {Wadhams}, \citenamefont {Bai},
		\citenamefont {Berry},\ and\ \citenamefont {Armitage}}]{Leake2006}%
	\BibitemOpen
	\bibfield  {author} {\bibinfo {author} {\bibfnamefont {M.\,C.}\ \bibnamefont
			{Leake}}, \bibinfo {author} {\bibfnamefont {J.\,H.}\ \bibnamefont {Chandler}},
		\bibinfo {author} {\bibfnamefont {G.\,H.}\ \bibnamefont {Wadhams}}, \bibinfo
		{author} {\bibfnamefont {F.}\,\bibnamefont {Bai}}, \bibinfo {author}
		{\bibfnamefont {R.\,M.}\ \bibnamefont {Berry}},\ and\ \bibinfo {author}
		{\bibfnamefont {J.\,P.}\ \bibnamefont {Armitage}},\ }\bibfield  {title}
	{\bibinfo {title} {{Stoichiometry and turnover in single, functioning
				membrane protein complexes}},\ }\href {https://doi.org/10.1038/nature05135}
	{\bibfield  {journal} {\bibinfo  {journal} {Nature}\ }\textbf {\bibinfo
			{volume} {443}},\ \bibinfo {pages} {355} (\bibinfo {year}
		{2006})}\BibitemShut {NoStop}%
	\bibitem [{\citenamefont {Lele}\ \emph {et\,al.}(2013)\citenamefont {Lele},
		\citenamefont {Hosu},\ and\ \citenamefont {Berg}}]{LeleHosuBerg2013}%
	\BibitemOpen
	\bibfield  {author} {\bibinfo {author} {\bibfnamefont {P.\,P.}\ \bibnamefont
			{Lele}}, \bibinfo {author} {\bibfnamefont {B.\,G.}\ \bibnamefont {Hosu}},\
		and\ \bibinfo {author} {\bibfnamefont {H.\,C.}\ \bibnamefont {Berg}},\
	}\bibfield  {title} {\bibinfo {title} {Dynamics of mechanosensing in the
			bacterial flagellar motor},\ }\href {https://doi.org/10.1073/pnas.1305885110}
	{\bibfield  {journal} {\bibinfo  {journal} {Proceedings of the National
				Academy of Sciences}\ }\textbf {\bibinfo {volume} {110}},\ \bibinfo {pages}
		{11839} (\bibinfo {year} {2013})}\BibitemShut {NoStop}%
	\bibitem [{\citenamefont {Tipping}\ \emph {et\,al.}(2013)\citenamefont
		{Tipping}, \citenamefont {Delalez}, \citenamefont {Lim}, \citenamefont
		{Berry},\ and\ \citenamefont
		{Armitage}}]{TippingDelalezLimBerryArmitage2013}%
	\BibitemOpen
	\bibfield  {author} {\bibinfo {author} {\bibfnamefont {M.\,J.}\ \bibnamefont
			{Tipping}}, \bibinfo {author} {\bibfnamefont {N.\,J.}\ \bibnamefont
			{Delalez}}, \bibinfo {author} {\bibfnamefont {R.}\,\bibnamefont {Lim}},
		\bibinfo {author} {\bibfnamefont {R.\,M.}\ \bibnamefont {Berry}},\ and\
		\bibinfo {author} {\bibfnamefont {J.\,P.}\ \bibnamefont {Armitage}},\
	}\bibfield  {title} {\bibinfo {title} {Load-dependent assembly of the
			bacterial flagellar motor},\ }\href {https://doi.org/10.1128/mbio.00551-13}
	{\bibfield  {journal} {\bibinfo  {journal} {mBio}\ }\textbf {\bibinfo
			{volume} {4}},\ \bibinfo {pages} {10.1128/mbio.00551} (\bibinfo {year}
		{2013})}\BibitemShut {NoStop}%
	\bibitem [{\citenamefont {Beeby}\ \emph {et\,al.}(2016)\citenamefont {Beeby},
		\citenamefont {Ribardo}, \citenamefont {Brennan}, \citenamefont {Ruby},
		\citenamefont {Jensen},\ and\ \citenamefont {Hendrixson}}]{Beeby2016}%
	\BibitemOpen
	\bibfield  {author} {\bibinfo {author} {\bibfnamefont {M.}\,\bibnamefont
			{Beeby}}, \bibinfo {author} {\bibfnamefont {D.\,A.}\ \bibnamefont {Ribardo}},
		\bibinfo {author} {\bibfnamefont {C.\,A.}\ \bibnamefont {Brennan}}, \bibinfo
		{author} {\bibfnamefont {E.\,G.}\ \bibnamefont {Ruby}}, \bibinfo {author}
		{\bibfnamefont {G.\,J.}\ \bibnamefont {Jensen}},\ and\ \bibinfo {author}
		{\bibfnamefont {D.\,R.}\ \bibnamefont {Hendrixson}},\ }\bibfield  {title}
	{\bibinfo {title} {{Diverse high-torque bacterial flagellar motors assemble
				wider stator rings using a conserved protein scaffold}},\ }\href
	{https://doi.org/10.1073/pnas.1518952113} {\bibfield  {journal} {\bibinfo
			{journal} {Proceedings of the National Academy of Sciences of the United
				States of America}\ }\textbf {\bibinfo {volume} {113}},\ \bibinfo {pages}
		{E1917} (\bibinfo {year} {2016})}\BibitemShut {NoStop}%
	\bibitem [{\citenamefont {Cohen}\ \emph {et\,al.}(2024)\citenamefont {Cohen},
		\citenamefont {Drobnic}, \citenamefont {Ribardo}, \citenamefont {Yoshioka},
		\citenamefont {Umrekar}, \citenamefont {Guo}, \citenamefont {Fernandez},
		\citenamefont {Brock}, \citenamefont {Wilson}, \citenamefont {Nakane},
		\citenamefont {Hendrixson},\ and\ \citenamefont {Beeby}}]{Cohen2024}%
	\BibitemOpen
	\bibfield  {author} {\bibinfo {author} {\bibfnamefont {E.}\,\bibnamefont
			{Cohen}}, \bibinfo {author} {\bibfnamefont {T.}\,\bibnamefont {Drobnic}},
		\bibinfo {author} {\bibfnamefont {D.}\,\bibnamefont {Ribardo}}, \bibinfo
		{author} {\bibfnamefont {A.}\,\bibnamefont {Yoshioka}}, \bibinfo {author}
		{\bibfnamefont {T.}\,\bibnamefont {Umrekar}}, \bibinfo {author} {\bibfnamefont
			{X.}\,\bibnamefont {Guo}}, \bibinfo {author} {\bibfnamefont {J.-J.}\
			\bibnamefont {Fernandez}}, \bibinfo {author} {\bibfnamefont {E.}\,\bibnamefont
			{Brock}}, \bibinfo {author} {\bibfnamefont {L.}\,\bibnamefont {Wilson}},
		\bibinfo {author} {\bibfnamefont {D.}\,\bibnamefont {Nakane}}, \bibinfo
		{author} {\bibfnamefont {D.\,R.}\ \bibnamefont {Hendrixson}},\ and\ \bibinfo
		{author} {\bibfnamefont {M.}\,\bibnamefont {Beeby}},\ }\href@noop {} {\bibinfo
		{title} {Evolution of a large periplasmic disk in {C}ampylobacterota flagella
			enables both efficient motility and autoagglutination}} (\bibinfo {year}
	{2024})\BibitemShut {NoStop}%
	\bibitem [{\citenamefont {Nord}\ \emph {et\,al.}(2016)\citenamefont {Nord},
		\citenamefont {Pedaci},\ and\ \citenamefont {Berry}}]{Nord2016}%
	\BibitemOpen
	\bibfield  {author} {\bibinfo {author} {\bibfnamefont {A.\,L.}\ \bibnamefont
			{Nord}}, \bibinfo {author} {\bibfnamefont {F.}\,\bibnamefont {Pedaci}},\ and\
		\bibinfo {author} {\bibfnamefont {R.\,M.}\ \bibnamefont {Berry}},\ }\bibfield
	{title} {\bibinfo {title} {Transient pauses of the bacterial flagellar motor
			at low load},\ }\href {https://doi.org/10.1088/1367-2630/18/11/115002}
	{\bibfield  {journal} {\bibinfo  {journal} {New Journal of Physics}\ }\textbf
		{\bibinfo {volume} {18}},\ \bibinfo {pages} {115002} (\bibinfo {year}
		{2016})}\BibitemShut {NoStop}%
	\bibitem [{\citenamefont {Lanoisel\'ee}\ \emph {et\,al.}(2025)\citenamefont
		{Lanoisel\'ee}, \citenamefont {Pagnini},\ and\ \citenamefont
		{Wylomanska}}]{LanoiseleePagniniWylomanska2025}%
	\BibitemOpen
	\bibfield  {author} {\bibinfo {author} {\bibfnamefont {Y.}\,\bibnamefont
			{Lanoisel\'ee}}, \bibinfo {author} {\bibfnamefont {G.}\,\bibnamefont
			{Pagnini}},\ and\ \bibinfo {author} {\bibfnamefont {A.}\,\bibnamefont
			{Wylomanska}},\ }\bibfield  {title} {\bibinfo {title} {Super-resolved
			anomalous diffusion: Deciphering the joint distribution of anomalous exponent
			and diffusion coefficient},\ }\href {https://doi.org/10.1103/y5pn-5ynd}
	{\bibfield  {journal} {\bibinfo  {journal} {Phys. Rev. Lett.}\ }\textbf
		{\bibinfo {volume} {135}},\ \bibinfo {pages} {137101} (\bibinfo {year}
		{2025})}\BibitemShut {NoStop}%
\end{thebibliography}

%

	\clearpage
	
	\section*{Supplementary information}
	\renewcommand{\thefigure}{S.\,\arabic{figure}} 
	\setcounter{figure}{0}
	\renewcommand{\thetable}{S.\,\Roman{table}} 
	\setcounter{table}{0}
	\subparagraph{Overdamped Langevin equation.}
	For Method 1, we demonstrate through several approaches, including numerical simulations of the Langevin equation, that for 
	$\epsilon<1$ and $\epsilon>1$ the dominant term in Eqs.\,(\ref{Eq:f/Kramers-1}) and (\ref{Eq:f/Pertur-2}), respectively, is $f_0 = \Theta\,\omega_{\rm ss}/D_{\rm ss}$.
	For the Smoluchowski equation, Eq.\,(\ref{Eq:Evol/Smoluchowski/rSpace}) with a 
	potential $V(\phi)=V_{\rm eq}(\phi)-f\phi$ the equivalent overdamped Langevin equation is:
	\begin{equation}\label{Eq:Evol/Langevin}\tag{SI.1}
		\frac{d \phi}{dt}=-\frac{1}{\gamma}\frac{\partial V(\phi)}{\partial\phi}+\xi(t)\,,
	\end{equation}
	where $\xi(t)$ is the thermal noise with zero mean and correlation $\langle \xi(t)\xi(t')\rangle=2\Theta\delta(t-t')/\gamma$. An ensemble of numerical simulations of Eq.\,(\ref{Eq:Evol/Langevin}) can be used to compute $P_{\rm ss}$, $\omega_{\rm ss}$ and $D_{\rm ss}$ in the same way as with real experimental data. The resulting simulations for a small enough time step converge to the analytic solution of the equivalent Smoluchowski description\,\cite{Gardiner2009}. 
	
\begin{figure}[htbp]
		\centering
		\includegraphics[height=175pt]{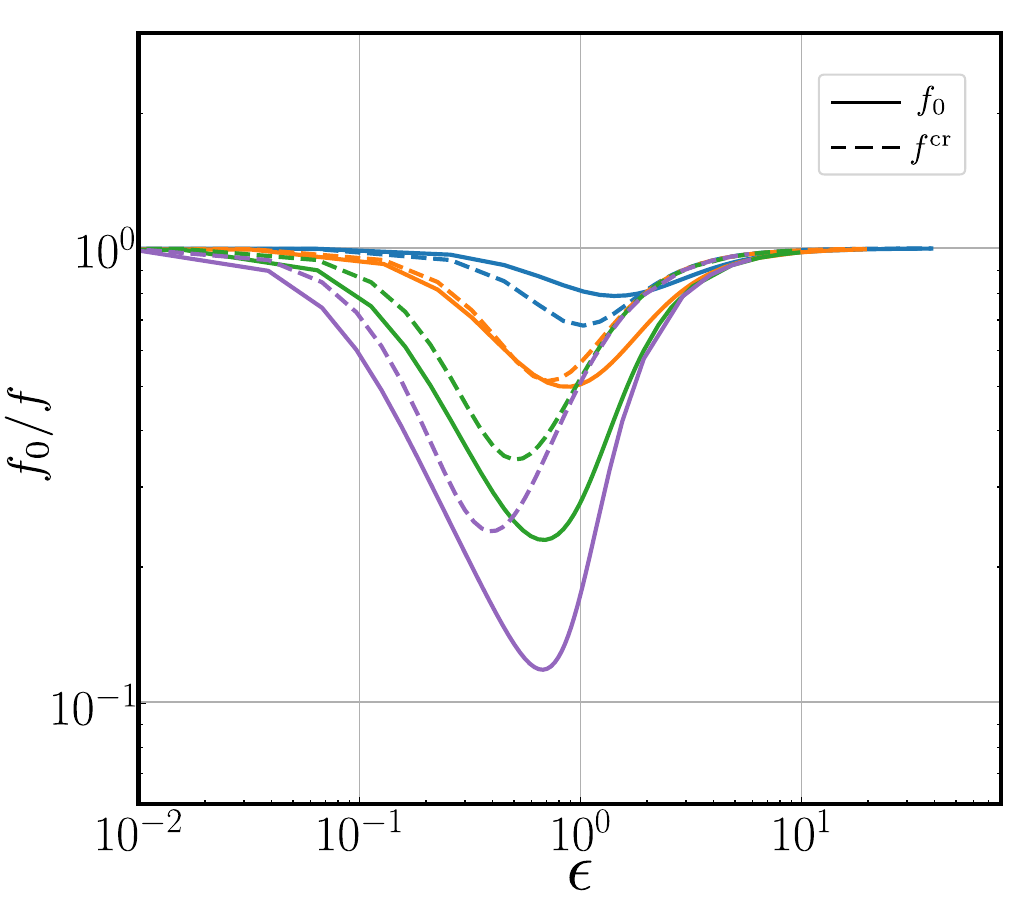}
		\caption{Relative error with $f$ of: $f_0$ (solid), Eq.\,(\ref{Eq:f/Padde}) $f^{\rm cr}$ (dashed). The different (online) colours represent increasing values of $E_{\rm a}$ are: (blue), (orange), (green), (purple)  $E_{\rm a}/\Theta=1.5, 3, 6, 10$ respectively.}
		\label{Fig:Pade_ss}
	\end{figure}

	\subparagraph{Pad\'{e} approximants.}
	The value of the tilt torque $f$ can be approximated by $f_0$ only in the close-to- and far-from-equilibrium regimes. In the cross-over region, the error incurred in this approximation is maximized, owing to the enhanced diffusion phenomenon. 
	It is known that the maximum of the diffusion is reached at the critical tilt $fL\simeq\pi E_{\rm a}$. In Figure\,\ref{Fig:Pade_ss} we show the relative error of $f_0$ against $f$ for increasing values of $E_{\rm a}$. 
	In addition, Figure\,\ref{Fig:Pade_ss} also shows the Pade approximant Eq.\,(\ref{Eq:f/Padde}) used to interpolate between the $\epsilon<1$ and $\epsilon>1$ regimes. It is worth noting that the Pade approximant is only a good approximation to the full $\Theta\omega_{\rm ss}/D_{\rm ss}$ dynamics for a restricted range of energy barriers, $E_{\rm a}\sim2$-4$\Theta$. 
	
\begin{figure}[htbp]
		\centering
		\includegraphics[height=175pt]{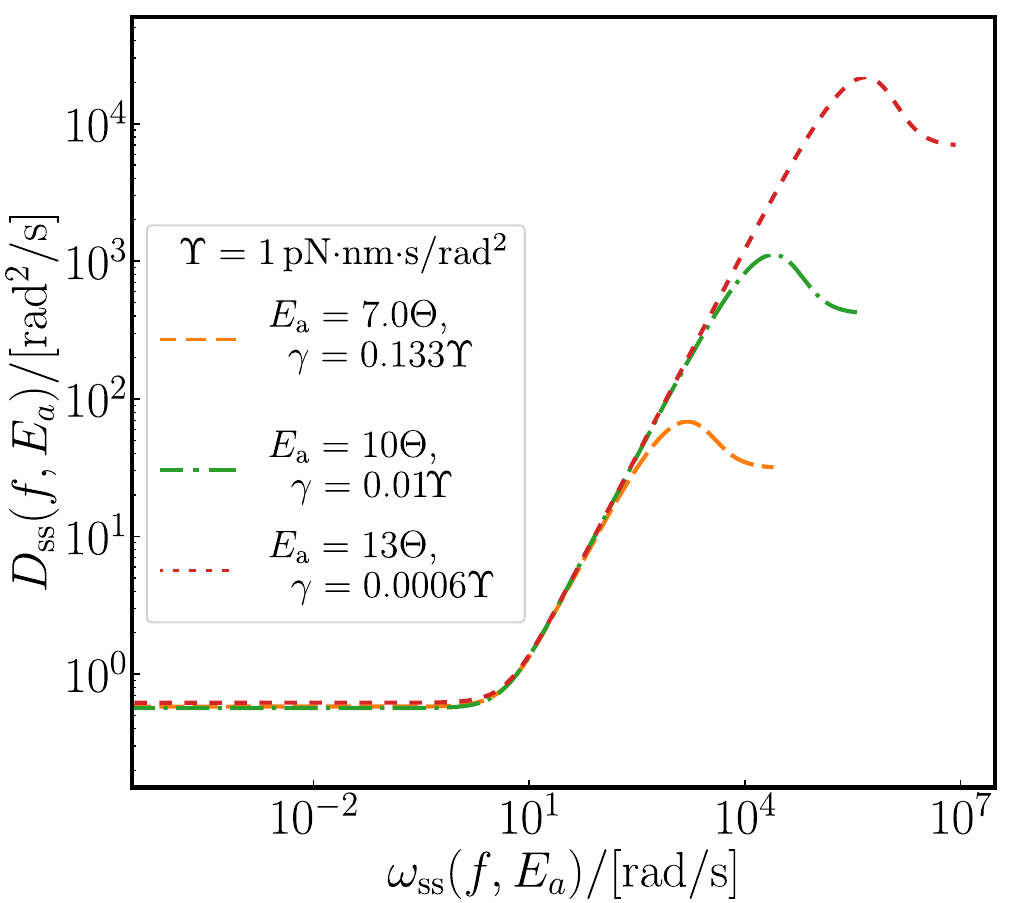}
		\caption{Comparative parametric plot of 
			$\{\omega_{\rm ss}, D_{\rm ss}\}$ for three given friction coefficient
			($\gamma$) values while fixing the 
			low-speed plateau. For the: (red), (green) and (orange) curves with 
			$\gamma$ values $0.0006, 0.01, 0.133$ \,\pDrg, the necessary 
			$E_{\rm a}/\Theta$ values to fix the low-speed plateau are 
			$13, 10, 7$, respectively.}
		\label{Fig:Same_Dk_diff_g}
	\end{figure}

	\subparagraph{Best fit to the 
		$\bm{D_{\rm ss}}$  vs. $\bm{\omega_{\rm ss}}$ data.} In the main text, by means of method 1 and the experimental dataset $\{\omega_{\rm ss}, D_{\rm ss}\}$, the best fit allowed us to determine $E_{\rm a}$, $L$ and $\gamma$. 
	Because of
	the lack of extensive data in the
	$\epsilon>1$ region, our estimate of $\gamma$ is an upper bound. If, instead, we imposed different values of 
	$\gamma$, while keeping constant the 
	$\epsilon<1$ plateau (see Fig.\,\ref{Fig:Same_Dk_diff_g} 
	low speed values), 
	we find a tradeoff between the $E_{\rm a}$ fitted value and the imposed $\gamma$ value (lower 
	$\gamma$ results in larger 
	$E_{\rm a}$ and the other way around). In Fig.\,\ref{Fig:Same_Dk_diff_g} we show an illustrative theoretical example of this tradeoff for three different 
	$\gamma$ values and their 
	corresponding
	$E_{\rm a}$ values. 
	
	\subparagraph{Complementary data.} The computed values for $\omega_{\rm ss}$, $D_{\rm ss}$ from each trace and their torque values according to the different expressions discussed in the main text are reported in Table\,\ref{Tab:TorqueRecons}.

\begin{figure}[htbp]
	\centering
	\includegraphics[height=175pt]{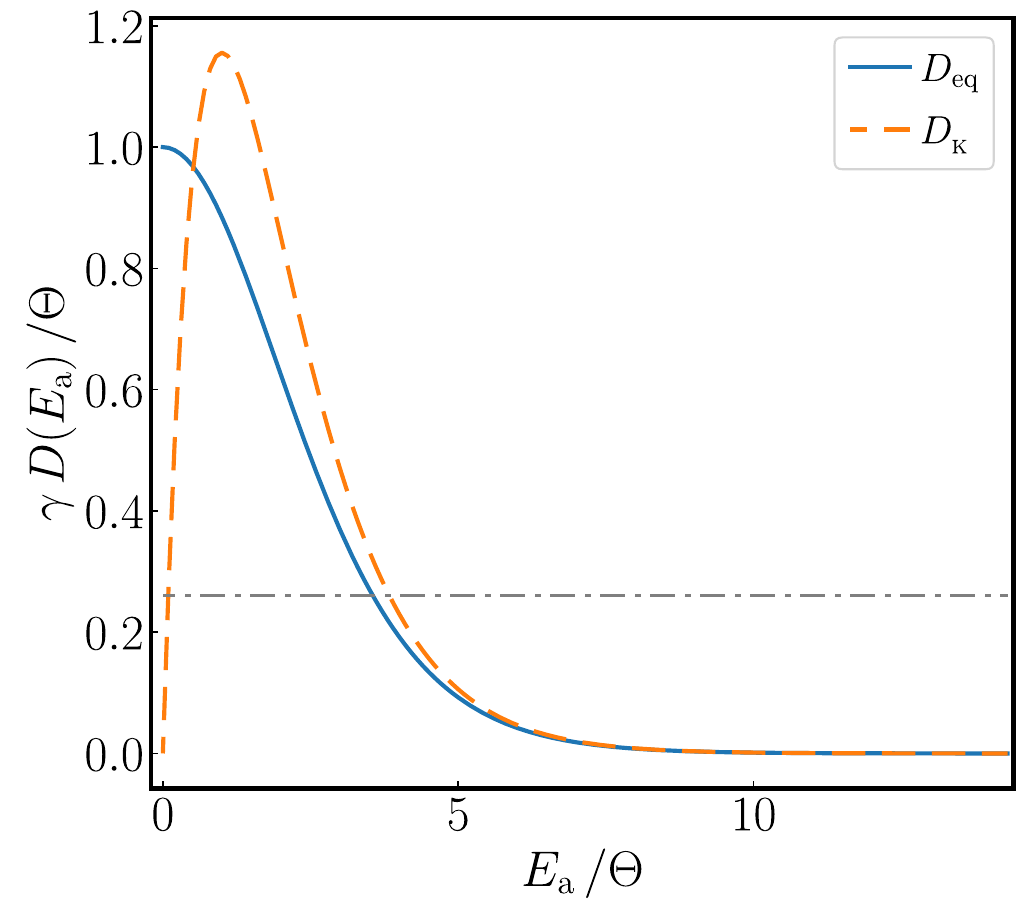}\put(-169,163){(a)}
	
	\includegraphics[height=175pt]{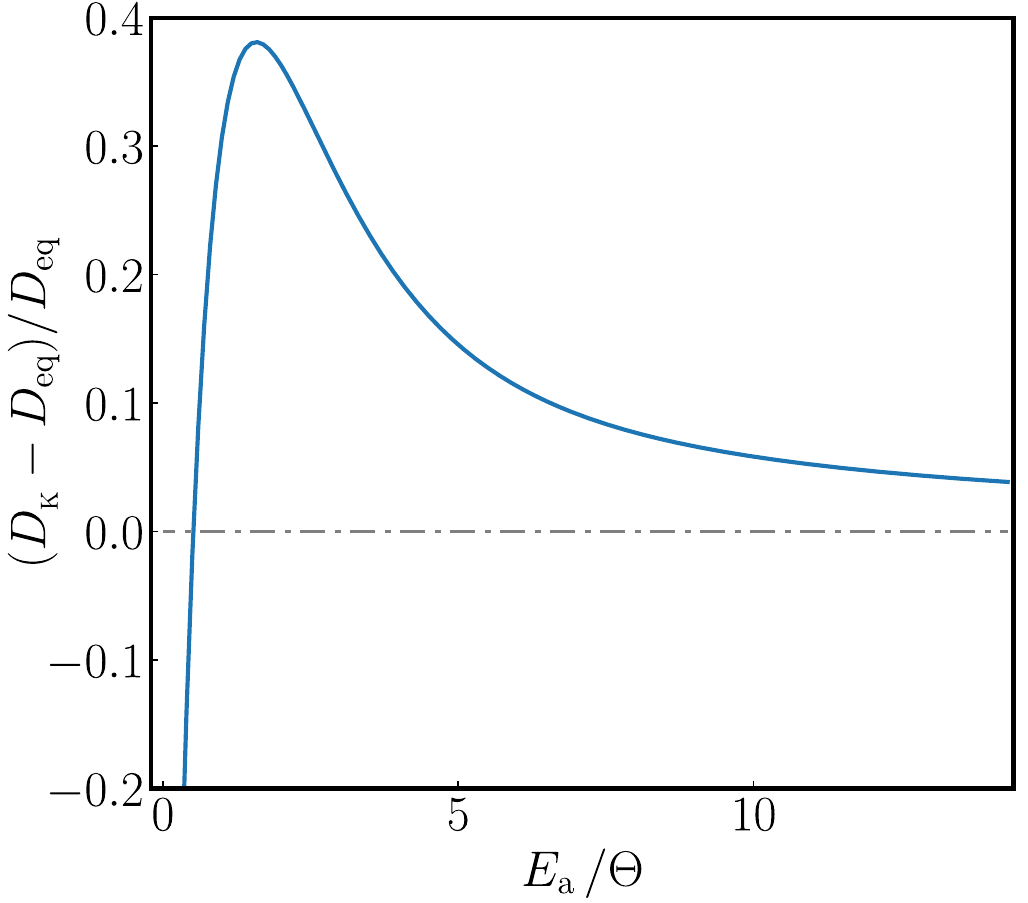}\put(-169,163){(b)}	
	\caption{
		(a) Kramers expression for the normalized equilibrium diffusion coefficient, $D_{\textsc{k}}/D_{\textsc{e}}$ (orange dashed curve) as a function of $E_{\rm a}/\Theta$;  normalized exact result $D_{\rm eq}/D_{\textsc{e}}$ Eq.\,SI.2 (blue solid curve); best model fit result (0.26) for the experimental equilibrium diffusion coefficient normalized by the free (Einstein) diffusion coefficient, $D_{\textsc{e}} =\Theta/\gamma$. 
		(b) The relative error incurred with respect to the exact result $D_{\rm eq}$ Eq.\,SI.2 (as a function of $E_{\rm a}/\Theta$) using the Kramers expression for the equilibrium diffusion coefficient, $D_{\textsc{k}}$ Eq.\,(\ref{Eq:DKramers}).}
	\label{Fig:RE_Exact_Kramers_Deq}
\end{figure}

	\subparagraph{Equilibrium diffusion beyond Kramers.} In the main text we introduced the  Kramers diffusion coefficient $D_{\textsc{k}}$, Eq.\,(\ref{Eq:DKramers}), which accounts for the suppression of diffusion by the landscape potential barriers at equilibrium and controls the asymptotic behavior at low speed of $D_{\rm ss}$ for a single-mode sinusoidal potential. This is a powerful, widely used approximation, 
	yet it can only be justified 
	for  barriers that are large enough compared to the thermal energy. The exact evaluation of Eqs.\,(\ref{Eq:vss/Reimann})-(\ref{Eq:Dss/Reimann}) gives the exact value at equilibrium
	\begin{equation}\tag{SI.2}
		D_{\rm eq}=\frac{\Theta}{\gamma}\bigg\vert I_{0}(E_{\rm a}/2\Theta)\bigg\vert^{-2},
	\end{equation}
	where $I_0$ is the modified zero order Bessel function of the 1st kind. 
	The Kramers expression overestimates $D_{\rm eq}$ for 
	$E_{\rm a}/\Theta > 0.5$. The relative error incurred is decreasing for $E_{\rm a}/\Theta > 1.6$ and less than 27\% for $E_{\rm a}/\Theta > 3$  (see Fig.\,\ref{Fig:RE_Exact_Kramers_Deq}).
	
	Unfortunately, no equivalent simple exact expression exists for diffusion in the case of more general potentials.
	
	\clearpage
	\onecolumngrid
	\begin{table*}[htb]
		\begin{minipage}[t]{0.45\textwidth}
			\centering
			\begin{tabular}{l>{\columncolor[gray]{0.87}}lc>{\columncolor[gray]{0.87}}cr>{\columncolor[gray]{0.87}}rr>{\columncolor[gray]{0.87}}r}
				\multicolumn{7}{c}{Data from Figure 2 Part 1}\\\cline{1-8}\\[-9pt]\cline{1-8}\\[-7pt]
				&\cellcolor{white}&             
				\cellcolor{white}&             
				\cellcolor{white}&             
				\cellcolor{white}&   
				\cellcolor{white}&             
				\cellcolor{white}&             
				\cellcolor{white}\\
				&\cellcolor{white}\multirow{-2}{*}{$\rm no.$}&                 
				\cellcolor{white}\multirow{-2}{*}{$\omega_{\rm ss}$}&   
				\cellcolor{white}\multirow{-2}{*}{$D_{\rm ss}$}&        
				\cellcolor{white}\multirow{-2}{*}{$f_\gamma^{\rm hi}\;\;$}&    
				\cellcolor{white}\multirow{-2}{*}{$f_0\;\;$}&       
				\cellcolor{white}\multirow{-2}{*}{$f_{\rm opt}\;\;$}&       
				\cellcolor{white}\multirow{-2}{*}{Eq.\,(\ref{Eq:vss/Reimann}):$f$}\\\cline{1-8}\\[-9pt]\cline{1-8}\\	
				
				&51 & 0.0239 & 3.3100 & 0.013 & 0.186 & 0.046 & 0.061 \\
				&53 & 0.2737 & 5.6800 & 0.150 & 1.241 & 0.526 & 0.697 \\
				&00 & 5.1035 & 15.8461 & 2.790 & 8.297 & 9.802 & 12.864 \\
				&03 & 6.2847 & 9.6883 & 3.435 & 16.711 & 8.137 & 15.758\\
				&01 & 7.5217 & 12.7418 & 4.112 & 15.207 & 14.446 & 18.738 \\
				&02 & 7.9044 & 13.8779 & 4.321 & 14.673 & 15.181 & 19.649 \\
				&04 & 8.7840 & 9.0493 & 4.802 & 25.006 & 16.870 & 21.719 \\
				&06 & 14.2429 & 21.3663 & 7.786 & 17.172 & 10.084 & 33.818 \\
				&05 & 17.0789 & 12.3978 & 9.336 & 35.488 & 13.420 & 39.575 \\
				&07 & 20.1734 & 32.1697 & 11.028 & 16.155 & 13.975 & 45.462 \\
				&08 & 21.2398 & 13.9638 & 11.610 & 39.184 & 21.907 & 47.400\\
				&10 & 21.6344 & 25.3848 & 11.826 & 21.955 & 20.223 & 48.107 \\
				&12 & 21.6743 & 15.4211 & 11.848 & 36.207 & 25.744 & 48.178 \\
				&11 & 22.4602 & 16.3930 & 12.278 & 35.295 & 25.084 & 49.565 \\
				&09 & 22.8635 & 15.7515 & 12.498 & 37.392 & 19.789 & 50.269 \\
				&13 & 23.0823 & 22.5076 & 12.618 & 26.419 & 28.640 & 50.648 \\
				&14 & 23.9548 & 23.7138 & 13.095 & 26.023 & 15.310 & 52.143 \\
				&15 & 25.0262 & 18.6213 & 13.680 & 34.622 & 15.426 & 53.941 \\
				&17 & 29.1267 & 27.1737 & 15.922 & 27.613 & 23.174 & 60.478 \\
				&20 & 29.8046 & 19.1550 & 16.292 & 40.083 & 27.798 & 61.510\\
				&16 & 30.4074 & 29.7798 & 16.622 & 26.304 & 19.390 & 62.417 \\
				&21 & 30.9447 & 20.2212 & 16.916 & 39.422 & 20.638 & 63.217 \\
				&19 & 32.3857 & 16.3387 & 17.703 & 51.062 & 28.930 & 65.324\\
				&23 & 32.8801 & 24.3842 & 17.974 & 34.737 & 26.389 & 66.036 \\
				&22 & 33.1172 & 24.0830 & 18.103 & 35.425 & 23.415 & 66.373\\
				&24 & 33.2831 & 24.6066 & 18.194 & 34.845 & 21.408 & 66.609\\
				&26 & 33.3958 & 22.1903 & 18.255 & 38.770 & 24.354 & 66.769 \\
				&18 & 33.4142 & 40.2038 & 18.265 & 21.410 & 23.053 & 66.796 \\
				&28 & 35.2927 & 21.5926 & 19.292 & 42.106 & 25.357 & 69.416 \\
				&27 & 35.7240 & 17.6973 & 19.528 & 52.002 & 23.887 & 70.008 \\
				&25 & 36.1628 & 29.7654 & 19.768 & 31.298 & 42.752 & 70.602 \\
				&31 & 36.6715 & 21.9235 & 20.046 & 43.091 & 31.138 & 71.288 \\
				&29 & 38.6531 & 28.1664 & 21.129 & 35.352 & 30.429 & 73.908 \\
				&32 & 39.7230 & 20.7943 & 21.714 & 49.211 & 27.245 & 75.287\\
				&30 & 41.1077 & 46.6582 & 22.471 & 22.696 & 37.327 & 77.042 \\
				&33 & 42.6200 & 21.8162 & 23.298 & 50.327 & 63.186 & 78.917\\
				&37 & 43.9138 & 30.7979 & 24.005 & 36.732 & 36.108 & 80.488 \\
				&35 & 43.9398 & 19.1909 & 24.019 & 58.983 & 30.668 & 80.519\\
				&34 & 45.8106 & 34.5847 & 25.042 & 34.123 & 37.653 & 82.743 \\
				&38 & 47.4519 & 25.7346 & 25.939 & 47.500 & 48.358 & 84.648 \\
				&36 & 47.5314 & 28.5718 & 25.982 & 42.855 & 46.077 & 84.739 \\
				&39 & 50.7467 & 24.5032 & 27.740 & $\cdots\;\;\;$ & 47.991 & 88.354 \\
				&40 & 53.4435 & 31.9690 & 29.214 & $\cdots\;\;\;$ & 64.140 & 91.280 \\
				&41 & 53.8306 & 50.6219 & 29.426 & $\cdots\;\;\;$ & 50.387 & 91.692\\
				&42 & 56.4306 & 46.7247 & 30.847 & $\cdots\;\;\;$ & 48.213 & 94.415 \\
				&43 & 63.3091 & 35.0726 & 34.607 & $\cdots\;\;\;$ & 79.035 & 101.270 \\
				&44 & 64.4417 & 53.1079 & 35.226 & $\cdots\;\;\;$ & 63.678 & 102.355\\
				&45 & 65.9242 & 29.5725 & 36.037 & $\cdots\;\;\;$ & 91.074 & 103.757 \\
				&47 & 66.8409 & 26.0788 & 36.538 & $\cdots\;\;\;$ & 110.320 & 104.615\\
			\end{tabular}
			
			\begin{tikzpicture}[remember picture, overlay]
				\draw[->, thick]
				([yshift=-3.7cm,xshift=-10.1cm]current page.north east) -- 
				([yshift=-20.3cm,xshift=-10.1cm]current page.north east)
				node[midway, rotate=90, anchor=center, yshift=8pt] {\small $\epsilon\simeq1$};
			\end{tikzpicture}
			
			\begin{tikzpicture}[remember picture, overlay]
				\draw[<-, thick]
				([yshift=-18.9cm,xshift=-19.3cm]current page.north east) -- 
				([yshift=-21.55cm,xshift=-19.3cm]current page.north east);
			\end{tikzpicture}

			\begin{tikzpicture}[remember picture, overlay]
				\draw[<->, thick]
				([yshift=-3.7cm,xshift=-19.3cm]current page.north east) -- 
				([yshift=-18.4cm,xshift=-19.3cm]current page.north east)
				node[midway, rotate=90, anchor=center, yshift=8pt] {\small $\epsilon<1$};
			\end{tikzpicture}
			
			\begin{tikzpicture}[remember picture, overlay]
				\draw[<->, thick]
				([yshift=-20.7cm,xshift=-10.1cm]current page.north east) -- 
				([yshift=-21.6cm,xshift=-10.1cm]current page.north east)
				node[midway, rotate=90, anchor=center, yshift=8pt] {\small $\epsilon>1$};
			\end{tikzpicture}
			
		\end{minipage}
		\hspace{0.05\textwidth} 
		\begin{minipage}[t]{0.45\textwidth}
			\centering
			\begin{tabular}{l>{\columncolor[gray]{0.87}}lc>{\columncolor[gray]{0.87}}cr>{\columncolor[gray]{0.87}}rr>{\columncolor[gray]{0.87}}r}
				\multicolumn{7}{c}{Data from Figure 2 Part 2}\\\cline{1-8}\\[-9pt]\cline{1-8}\\[-7pt]
				&\cellcolor{white}&             
				\cellcolor{white}&             
				\cellcolor{white}&             
				\cellcolor{white}&   
				\cellcolor{white}&             
				\cellcolor{white}&             
				\cellcolor{white}\\
				&\cellcolor{white}\multirow{-2}{*}{$\rm no.$}&                 
				\cellcolor{white}\multirow{-2}{*}{$\omega_{\rm ss}$}&   
				\cellcolor{white}\multirow{-2}{*}{$D_{\rm ss}$}&        
				\cellcolor{white}\multirow{-2}{*}{$f_\gamma^{\rm hi}\;\;$}&    
				\cellcolor{white}\multirow{-2}{*}{$f_0\;\;$}&       
				\cellcolor{white}\multirow{-2}{*}{$f_{\rm opt}\;\;$}&       
				\cellcolor{white}\multirow{-2}{*}{Eq.\,(\ref{Eq:vss/Reimann}):$f$}\\\cline{1-8}\\[-9pt]\cline{1-8}\\	
				
				&46 & 70.1835 & 47.2294 & 38.365 & $\cdots\;\;\;\;$  & 42.077 & 107.684 \\
				&48 & 74.1483 & 46.6297 & 40.532 & $\cdots\;\;\;\;$ & 78.975 & 111.215 \\
				&49 & 87.7138 & 100.2180 & 47.948 & $\cdots\;\;\;\;$  & 53.513 & 122.525 \\
				&52 & 94.3950 & 54.7888 & 51.600 & $\cdots\;\;\;\;$  & 56.238 & 127.732 \\
				&51* & 94.5640 & 30.4585 & 51.692 & $\cdots\;\;\;\;$  & 89.330 & 127.861 \\
				&50 & 95.4453 & 37.2578 & 52.174 & $\cdots\;\;\;\;$  & 56.210 & 128.531 \\
				&53* & 99.9322 & 48.6768 & 54.627 & $\cdots\;\;\;\;$  & 109.323 & 131.896 \\
				&54 & 100.3669 & 61.0488 & 54.864 & $\cdots\;\;\;\;$  & 59.157 & 132.218 \\
				&56 & 103.1939 & 74.0296 & 56.410 & $\cdots\;\;\;\;$  & 62.059 & 134.291\\
				&57 & 105.2385 & 32.7892 & 57.527 & $\cdots\;\;\;\;$  & 72.303 & 135.772 \\
				&55 & 107.4102 & 83.7385 & 58.714 & $\cdots\;\;\;\;$  & 79.657 & 137.329\\
				&67 & 107.8466 & 54.5860 & 58.953 & $\cdots\;\;\;\;$  & 74.386 & 137.640 \\
				&65 & 108.2996 & 51.3934 & 59.201 & $\cdots\;\;\;\;$  & 72.201 & 137.962 \\
				&59 & 108.5291 & 47.8969 & 59.326 & $\cdots\;\;\;\;$  & 64.579 & 138.125 \\
				&58 & 109.7325 & 50.0629 & 59.984 & $\cdots\;\;\;\;$  & 107.263 & 138.977 \\
				&63 & 109.9765 & 49.3522 & 60.117 & $\cdots\;\;\;\;$  & 86.300 & 139.148\\
				&60 & 110.9838 & 39.8473 & 60.668 & $\cdots\;\;\;\;$  & 108.999 & 139.856 \\
				&64 & 112.6764 & 82.1488 & 61.593 & $\cdots\;\;\;\;$  & 128.716 & 141.039 \\
				&69 & 115.2528 & 39.1144 & 63.001 & $\cdots\;\;\;\;$  & 82.014 & 142.821 \\
				&61 & 115.4305 & 84.7919 & 63.099 & $\cdots\;\;\;\;$  & 97.473 & 142.943 \\
				&62 & 120.5208 & 108.8620 & 65.881 & $\cdots\;\;\;\;$  & 74.870 & 146.404 \\
				&66 & 121.8290 & 73.8481 & 66.596 & $\cdots\;\;\;\;$  & 75.381 & 147.282 \\
				&68 & 122.3359 & 81.1375 & 66.873 & $\cdots\;\;\;\;$  & 71.569 & 147.621 \\
				&70 & 126.6816 & 42.4942 & 69.249 & $\cdots\;\;\;\;$  & 72.933 & 150.50 \\
				&72 & 127.9478 & 56.4552 & 69.941 & $\cdots\;\;\;\;$  & 74.182 & 151.330 \\
				&71 & 131.9177 & 55.9759 & 72.111 & $\cdots\;\;\;\;$  & 78.451 & 153.907 \\
				&75 & 146.2412 & 57.0763 & 79.941 & $\cdots\;\;\;\;$  & 116.777 & 162.923 \\
				&74 & 146.3928 & 66.5373 & 80.024 & $\cdots\;\;\;\;$  & 114.154 & 163.017 \\
				&73 & 148.1676 & 64.5067 & 80.994 & $\cdots\;\;\;\;$  & 88.313 & 164.106\\
				&76 & 148.9227 & 67.4750 & 81.407 & $\cdots\;\;\;\;$  & 119.517 & 164.567 \\
				&77 & 148.9489 & 60.2831 & 81.421 & $\cdots\;\;\;\;$  & 90.287 & 164.583 \\
				&79 & 156.9166 & 74.8191 & 85.776 & $\cdots\;\;\;\;$  & 117.516 & 169.395 \\
				&78 & 162.2314 & 55.1422 & 88.682 & $\cdots\;\;\;\;$  & 99.004 & 172.549 \\
				&80 & 166.6352 & 54.7579 & 91.089 & $\cdots\;\;\;\;$  & 103.740 & 175.131 \\
				&81 & 167.8469 & 51.6078 & 91.751 & $\cdots\;\;\;\;$  & 96.954 & 175.837 \\
				&82 & 181.5321 & 111.1560 & 99.232 & $\cdots\;\;\;\;$  & 105.557 & 183.678\\
				&83 & 188.0893 & 88.4304 & 102.817 & $\cdots\;\;\;\;$  & 110.620 & 187.359 \\
				&84 & 190.2920 & 68.5888 & 104.021 & $\cdots\;\;\;\;$  & 110.263 & 188.585\\
				&86 & 209.7508 & 84.4314 & 114.658 & $\cdots\;\;\;\;$  & 126.207 & 199.223 \\
				&85 & 210.8596 & 67.2377 & 115.264 & $\cdots\;\;\;\;$  & 128.694 & 199.819 \\
				&87 & 215.3902 & 94.9795 & 117.740 & $\cdots\;\;\;\;$  & 132.658 & 202.247 \\
				&88 & 221.6105 & 65.0147 & 121.141 & $\cdots\;\;\;\;$  & 132.684 & 205.556\\
				&89 & 223.2779 & 88.1435 & 122.052 & $\cdots\;\;\;\;$  & 132.615 & 206.439\\
				&90 & 277.7583 & 92.3734 & 151.833 & $\cdots\;\;\;\;$  & 180.393 & 234.447 \\
				&91 & 302.2618 & 75.4917 & 165.228 & $\cdots\;\;\;\;$  & 176.938 & 246.666 \\
				&92 & 352.6364 & 96.1755 & 192.764 & $\cdots\;\;\;\;$  & 205.022 & 271.361 \\
				&93 & 357.3548 & 87.2789 & 195.343 & 105.476 & 216.975 & 273.655\\
				&94 & 416.5345 & 69.9834 & 227.693 & 153.327 & 248.986 & 302.271 \\
				&95 & 483.4029 & 69.9267 & 264.246 & 178.086 & 338.451 & 334.510 \\
			\end{tabular}
		\end{minipage} 
		\caption{Experimental traces with calculated observables $\omega_{\rm ss}$ [Hz], $D_{\rm ss}$ [rad$^2$/s], and estimated torque [pN$\cdot$nm/rad] via: (4th col) $f_\gamma^{\rm hi}$ Eq.\,(\ref{Eq:f/gamma}), (5th col) $f_0$ \textbf{method 1}, (6th col) $f_{\rm opt}$ \textbf{method 3}, and (7th col) Eq.\,(\ref{Eq:vss/Reimann}) theoretical prediction with $E_{\rm a}=3.8\Theta$. (*) Observables calculated at a different time interval than table part 1. ($\;\cdots\;$) 
			In accordance 
			with our findings $E_{\rm a}\sim2$-4$\Theta$,  any trace with $350\,\text{Hz}\gtrsim\omega_{\rm ss}\gtrsim50\,\text{Hz}$  is outside the validity range for $f_0$. For both $f_\gamma^{\rm hi}$ Eq.\,(\ref{Eq:f/gamma}) and  Eq.\,(\ref{Eq:vss/Reimann}) we used $\gamma=0.1$\,\pDrg.}\label{Tab:TorqueRecons}
	\end{table*}

\end{document}